\documentclass[12pt,letterpaper]{article}
\usepackage{epsfig,rotating,setspace,latexsym,amsmath,epsf,amssymb,amsfonts,bm,theorem,cite,caption,subcaption,enumerate,longtable,accents}
\usepackage{algorithm,algorithmic,graphicx,epsf,authblk,epstopdf,url,color,multirow}

\newtheorem{theorem}{Theorem}

\newtheorem{corollary}{Corollary}

\newtheorem{remark}{Remark}

\newenvironment{Proof}[1]{\medskip\par\noindent{\bf Proof:\,}\,#1}{{\mbox{\,$\blacksquare$}\par}}

\allowdisplaybreaks

\begin{document}

\title{Semantic Private Information Retrieval\thanks{This work was supported by NSF Grants CCF 17-13977 and ECCS 18-07348.}}

\author[1]{Sajani Vithana}
\author[2]{Karim Banawan}
\author[1]{Sennur Ulukus}
\affil[1]{\normalsize Department of Electrical and Computer Engineering, University of Maryland}
\affil[2]{\normalsize Electrical Engineering Department, Faculty of Engineering, Alexandria University}
 
\maketitle

\vspace*{-1.0cm}

\begin{abstract}
We investigate the problem of semantic private information retrieval (semantic PIR). In semantic PIR, a user retrieves a message out of $K$ independent messages stored in $N$ replicated and non-colluding databases without revealing the identity of the desired message to any individual database. The messages come with \emph{different semantics}, i.e., the messages are allowed to have \emph{non-uniform a priori probabilities} denoted by $(p_i>0,\: i \in [K])$, which are a proxy for their respective popularity of retrieval, and \emph{arbitrary message sizes} $(L_i,\: i \in [K])$. This is a generalization of the classical private information retrieval (PIR) problem, where messages are assumed to have equal a priori probabilities and equal message sizes. We derive the semantic PIR capacity for general $K$, $N$. The results show that the semantic PIR capacity depends on the number of databases $N$, the number of messages $K$, the a priori probability distribution of messages $p_i$, and the message sizes $L_i$. We present two achievable semantic PIR schemes: The first one is a deterministic scheme which is based on message asymmetry. This scheme employs non-uniform subpacketization. The second scheme is probabilistic and is based on choosing one query set out of multiple options at random to retrieve the required message without the need for exponential subpacketization. We derive necessary and sufficient conditions for the semantic PIR capacity to exceed the classical PIR capacity with equal priors and sizes. Our results show that the semantic PIR capacity can be larger than the classical PIR capacity when longer messages have higher popularities. However, when messages are equal-length, the non-uniform priors cannot be exploited to improve the retrieval rate over the classical PIR capacity.   
\end{abstract}

\section{Introduction}

Private information retrieval (PIR) describes an elemental privacy setting. In the classical PIR problem, introduced in the seminal paper \cite{PIR_ORI}, a user needs to retrieve a message (file), out of several messages, from multiple replicated databases, without revealing any information about the identity of the desired message. This problem has attracted significant recent interest in information theory where the fundamental limits of the problem based on absolute guarantees (in contrast to computational guarantees as in \cite{yekhanin2010private}) have been investigated. In \cite{sun2017capacity}, the notion of PIR capacity is introduced as the maximum ratio of the desired message size to the total download size. Reference \cite{sun2017capacity} has characterized the classical PIR capacity using a greedy algorithm which is based on message and database symmetry. Using this performance metric, further practical variants of the problem have been investigated \cite{JafarColluding, arbitraryCollusion, disjointJafar, arbitrary-coll-Kang, Staircase_PIR, codedsymmetric, SPIR_Mismatched, SPIR,  ChaoTian_leakage, KarimCoded, codedcolluded, codedcolludingZhang, Kumar_PIRarbCoded, codedcolludingJafar, MM-PIR, MPIRcodedcolludingZhang, BPIRjournal, CodeColludeByzantinePIR, Byzantine-Kang,tandon2017capacity, KimCache, wei2017fundamental, wei2017fundamental_partial, PIR_cache_edge, kadhe2017private, chen2017capacity, wei2017capacity, MMPIR_PSI, SSMMPIR_SI1, SSMMPIR_SI2, LiConverse, StorageConstrainedPIR_Wei, PrivateComputation, mirmohseni2017private, PrivateSearch, abdul2017private, StorageConstrainedPIR, PIR_decentralized, heteroPIR, efficient_storage_ITW2019, Chao_storage_cost, TamoISIT, Karim_nonreplicated, SecurePIR, securePIRcapacity, PIR_WTC_II, securestoragePIR,  XSTPIR, arbmsgPIR, ChaoTian_coded_minsize, MultiroundPIR, KarimAsymmetricPIR, noisyPIR, PIR_lifting, PIR_networks, PSIjournal, PIRlatent}.  

In all these works, two assumptions are made: All messages have the same size\footnote{With the exception of \cite{SPIR}, which characterizes the capacity of the symmetric PIR (SPIR) problem for heterogeneous file sizes (without considering a priori probabilities of retrieval) to be $R_k=\frac{L_k}{\max_i \: L_i}\left(1-\frac{1}{N}\right)$, where $R_k$ is the rate of retrieving message $k$. The achievable scheme follows by dividing the files into partitions of length $N-1$ and repeating the original SPIR scheme in each partition. This scheme zero-pads shorter messages so that their lengths are equal to that of the longest message.}, $L$, and all messages are requested uniformly by the users. These assumptions are highly idealistic from a practical point of view. Take a streaming application for instance. The storage database has a catalog of different movies and TV shows. These media files cannot be assumed to have the same level of popularity, i.e., it is unlikely that all files are equally probable to be downloaded by a user. The streaming service, in this case, has an a priori probability distribution over all the files, for example, from box office revenues and online rating systems. In addition, the media files cannot be assumed to be equal in size; some movies are longer, some are shorter. Consequently, each message stored in the databases exhibits different \emph{semantics}, in the sense that each message has a different size and a different prior probability of retrieval. With this backdrop, in this paper, we investigate how a PIR scheme should be implemented over databases holding messages with different semantics. 

In this paper, we introduce the semantic PIR problem. We extend the notion of the PIR capacity to deal with the heterogeneity of message sizes and prior probabilities. We define the retrieval rate to be the ratio of the \emph{expected} message size to the \emph{expected} download cost. Due to the privacy constraint, the download cost needs to be the same for all messages; thus, the expected download cost is equal to the download cost for each individual message. Hence, the retrieval rate achieved by a given scheme is equal to the weighted average of all individual message retrieval rates. We investigate the semantic PIR capacity as a function of the system parameters: number of databases $N$, number of messages $K$, message priors $p_i$, and message lengths $L_i$. We ask how semantic PIR capacity compares to classical PIR capacity, and whether there is a PIR capacity gain from exploiting the message semantics. 

In this paper, we characterize the exact semantic PIR capacity for arbitrary parameters. To that end, we present two achievable schemes; the first scheme is deterministic, in the sense that the \emph{query structure} is fixed, and the second scheme is stochastic, in the sense that the user picks a query structure \emph{randomly} from a list of possible structures. For the deterministic scheme, we present a systematic method to determine the subpacketization level for each message. Note that this is crucial in our semantic problem due to the heterogeneous message sizes, unlike the majority of the literature that utilizes uniform subpacketization within their schemes \cite{8437880}. This scheme uses non-uniform subpacketization where the block size considered in each download differs from one message to another. The query structure of the deterministic scheme resembles the query structure of \cite{sun2017capacity}, in that, our scheme uses the same $k$-sums idea of \cite{sun2017capacity}. The second achievable scheme is comprised of several query options that the user may use with equal probability to retrieve any message. In this scheme, the messages are divided into several blocks depending on the number of databases. The message is retrieved using a single set of queries, which is chosen uniformly randomly from the query options to ensure privacy. This is similar to the scheme presented in \cite{ravi-leaky} with an extension to more than two databases (see also \cite{chao-tian}). We provide a matching converse that takes into account the heterogeneity of message sizes, resulting in settling the semantic PIR capacity. 

The semantic PIR capacity is a function of the message sizes and the a priori probability distribution. The expression implies that for certain message sizes and priors, the classical PIR capacity may be exceeded by exploiting the semantics of the messages even if the zero-padding needed in classical PIR to equalize the message sizes is ignored. Concretely, our results imply: 1) When message lengths are the same, semantic PIR capacity is equal to the classical PIR capacity no matter what the message priors are, i.e., priors cannot be exploited to increase the PIR capacity if the message lengths are the same. 2) For certain cases, such as when the prior probability distribution favors longer files (i.e., longer files are more popular), the semantic PIR capacity exceeds the classical PIR capacity which depends only on the number of databases and the number of messages. Note that, by classical PIR capacity, we mean the classical PIR capacity expression, which may not be attainable for heterogeneous file sizes. 3)  For all priors and lengths, our scheme achieves a larger PIR rate than the PIR rate the classical approach would achieve by simply zero-padding the messages to bring them to the same length, as it assumes. 

\section{Problem Formulation}\label{SPIR-formulation}
We consider a setting, where $N$ non-colluding databases store $K$ independent messages (files), $W_1,\dots,W_K$, in a replicated fashion. The messages exhibit different semantics, i.e., the messages have different sizes and different a priori probabilities of retrieval. The a priori probability of $W_i$ is denoted by\footnote{We assume that $p_i>0$ for all $i \in [K]$ without loss of generality, as $p_j=0$ for some $j$ implies that this message, $W_j$, is either non-existent or never requested by the user. Hence, the setting can be reduced to a semantic PIR problem with $K-1$ messages, each with $p_i>0$.} $p_i$, such that $p_i>0$ for $i=1,\dots,K$. The a priori probability distribution is globally known at the databases and the user. We assume that all message symbols are picked from a finite field\footnote{In this work, it suffices to work with the binary field, hence, symbols can be interpreted as bits.} $\mathbb{F}_s$. The message size of the $i$th message is denoted by $L_i$. Without loss of generality, we assume that the messages are ordered with respect to their sizes\footnote{This is for ease of expression of the capacity formula in (\ref{semantic_capacity}). The largest length should have the largest coefficient in the expression in (\ref{semantic_capacity}) in order to have the largest achievable rate and the tightest converse.}, such that  $L_1 \geq L_2 \geq \dots \geq L_K$. Since the messages stored in databases are independent of each other, the mutual information between them is zero, 
\begin{align}
    I(W_i;W_j)=&0, \quad  i\neq j
\end{align}
The message sizes can be expressed in $s$-ary symbols as,
\begin{align}
    H(W_i)&=L_i, \quad i=1, \dots, K \\
    H(W_1, \dots, W_K)&=\sum_{i=1}^K H(W_i)=\sum_{i=1}^K L_i
\end{align}

In semantic PIR, a user needs to retrieve a message $W_i$ without revealing the index $i$ to any individual database. To that end, the user sends a query to each database. The query sent to the $n$th database to retrieve $W_i$ is denoted by $Q_n^{[i]}$ for $n=1, \dots, N$. Prior to retrieval, the user does not have any information about the message contents. Hence, queries sent to the databases to retrieve messages are independent of the messages, i.e., the mutual information between messages and queries is zero,
\begin{align}
    I(W_1,\dots,W_K ; Q^{[i]}_1,\dots,Q^{[i]}_N)=0,\quad i=1,\dots,K
\end{align}

Once the databases receive the queries, they generate answer strings to send back to the user. Specifically, the $n$th database prepares an answer string $A_n^{[i]}$ which is a deterministic function of the stored messages $W_1,\dots,W_K$ and the received query $Q_n^{[i]}$. Therefore,
\begin{align}
    H(A^{[i]}_n|Q^{[i]}_n,W_1,\dots,W_K)=0,\quad i=1,\dots,K, \quad n=1,\dots,N
\end{align}

For a feasible PIR scheme, two conditions need to be satisfied, namely, the correctness and the privacy constraints. These are formally described as follows.

\textbf{Correctness:} The user should be able to perfectly retrieve the desired message as soon as the answer strings to the queries are received from the respective databases. Therefore,
\begin{align}\label{correctness}
    H(W_i|A^{[i]}_1,\dots,A^{[i]}_N,Q^{[i]}_1,\dots,Q^{[i]}_N)=0,\quad i=1,\dots,K
\end{align}

\textbf{Privacy:} To protect the privacy of the desired message index $i$, the queries should not leak any information about $i$. Formally, for the $n$th database, the a posteriori probability of the message index $i$ given a query $Q_n^{[i]}$ should be equal to the a priori probability of the message index $i$. That is, the random variable representing the desired message index, $\theta$, should be independent of the received set of queries. Therefore,
\begin{align}\label{privacy1}
    P(\theta=i|Q_n^{[i]})=P(\theta=i), \quad i=1,\dots,K, \quad n=1, \dots, N
\end{align}
The privacy constraint \eqref{privacy1} implies,
\begin{align}
    (Q^{[i]}_n,A^{[i]}_n,W_1,\dots,W_K)\sim(Q^{[j]}_n,A^{[j]}_n,W_1,\dots,W_K), ~ n=1,\dots,N, ~ i,j=1,\dots,K, ~ i\neq j
\end{align}

An achievable semantic PIR scheme is a scheme that satisfies the correctness constraint \eqref{correctness} and the privacy constraint \eqref{privacy1}. Due to the heterogeneity of message sizes and a priori probabilities, in this work, we define the performance metric, the expected retrieval rate $R$, as the ratio of the expected retrieved message size to the expected download size, i.e., 
\begin{align} \label{ach-rate}
    R=\frac{\mathbb{E}[L]}{\mathbb{E}[D]}
\end{align}
where $\mathbb{E}[L]$ is the expected number of useful bits downloaded and $\mathbb{E}[D]$ is the expected number of total bits downloaded. The expectation $\mathbb{E}[\cdot]$ is with respect to the a priori probability distribution. The semantic PIR capacity is defined as the supremum of the expected retrieval rates over all achievable PIR schemes, i.e., $C=\sup \: R$.

\section{Main Results and Discussions}
In this section, we present the main results of the paper. Our first result is a complete characterization of the semantic PIR capacity. The semantic PIR capacity depends on the message sizes and prior probability distribution. 

\begin{theorem}\label{Thm2}
The semantic PIR capacity with $N$ databases, $K$ messages, message sizes $L_i$ (arranged in decreasing order as $L_1 \geq L_2 \geq \dots \geq L_K$), and prior probabilities $p_i$, is
\begin{align}\label{semantic_capacity}
    C&= \left(\frac{L_1}{\mathbb{E}[L]}+\frac{1}{N}\frac{L_2}{\mathbb{E}[L]}+\dots+\frac{1}{N^{K-1}}\frac{L_K}{\mathbb{E}[L]}\right)^{-1}\\
     &=\left(\frac{L_1}{\sum_{i=1}^K p_i L_i}+\frac{1}{N}\frac{L_2}{\sum_{i=1}^K p_i L_i}+\dots+\frac{1}{N^{K-1}}\frac{L_K}{\sum_{i=1}^K p_i L_i}\right)^{-1}
\end{align}
where $\mathbb{E}[L]=\sum_{i=1}^K p_i L_i$.
\end{theorem}

The achievability proof of Theorem~\ref{Thm2} is presented in Section~\ref{Achievability} and the converse proof is presented in Section~\ref{converse}. Next, we have a few corollaries and remarks.

The following corollary gives a necessary and sufficient condition for the cases at which the semantic capacity exceeds the classical PIR capacity.

\begin{corollary}[A Necessary and Sufficient Condition for Semantic Capacity Gain]
The semantic PIR capacity is strictly larger than the classical PIR capacity (with uniform priors and message sizes) if and only if,
\begin{align}\label{corollary1}
    \sum_{i=1}^K \frac{1}{N^{i-1}} (L_i-\mathbb{E}[L])<0
\end{align}
which is further equivalent to,
\begin{align}
    \sum_{i=1}^K \sum_{j=1}^K \frac{p_j}{N^{i-1}}(L_i-L_j)<0
\end{align}
\end{corollary}

\begin{Proof}
The proof follows from comparing the semantic PIR capacity expression in \eqref{semantic_capacity} and the classical PIR capacity, $C_{PIR}$, in \cite{sun2017capacity},
\begin{align} \label{C_PIR}
    C_{PIR}=\left(1+\frac{1}{N}+\dots+\frac{1}{N^{K-1}}\right)^{-1}
\end{align}
Hence, $C>C_{PIR}$ implies
\begin{align}
    \frac{L_1}{\mathbb{E}[L]}+\frac{1}{N}\frac{L_2}{\mathbb{E}[L]}+\dots+\frac{1}{N^{K-1}}\frac{L_K}{\mathbb{E}[L]}<1+\frac{1}{N}+\dots+\frac{1}{N^{K-1}}
\end{align}
Ordering the terms leads to,
\begin{align}
        \sum_{i=1}^K \frac{1}{N^{i-1}} (L_i-\mathbb{E}[L])<0
\end{align}
Noting $L_i=\sum_{j=1}^K p_j L_i$, since $p_j$ sum to $1$, and $\mathbb{E}[L]=\sum_{j=1}^K p_j L_j$ by definition of expectation,
\begin{align}
    \sum_{i=1}^K \sum_{j=1}^K \frac{p_j}{N^{i-1}}(L_i-L_j)<0
\end{align}
\end{Proof}

\begin{remark}
The condition in \eqref{corollary1} is a statement about the sum weighted (by $\frac{1}{N^{i-1}}$) deviation of message size from its expected value. Note that the expected value of the message size $\mathbb{E}[L]$ is a function of the message sizes $L_i$ and the prior distribution $p_i$ for $i=1, \dots, K$.
\end{remark}

\begin{remark} \label{remark2}
More explicit conditions can be derived for specific cases. For example, consider the case $K=2$, $N=2$, and assume that $L_1 >L_2$ (strictly larger). Then, \eqref{corollary1} simplifies to,
\begin{align}
    (L_1-(p_1 L_1+p_2 L_2)) + \frac{1}{2}(L_2-(p_1 L_1+p_2L_2))&<0\\
    p_2(L_1-L_2)+\frac{1}{2}p_1(L_2-L_1)&<0\\
    p_2-\frac{1}{2}p_1&<0\label{remark2-eqn}\\
    p_1&>\frac{2}{3}
\end{align}
where \eqref{remark2-eqn} follows from $L_1>L_2$. This means that for $N=2$ and $K=2$, the capacity of semantic PIR is greater than the capacity of classical PIR when the a priori probability of the longer message is greater than $\frac{2}{3}$ irrespective of the values of $L_1$ and $L_2$.

As a further explicit example, if the more likely message is $4$ times more likely and $4$ times longer than the less likely message, i.e., if $p_1=4 p_2$ and $L_1=4 L_2$, then the semantic PIR capacity is $C=\frac{34}{45}$ while the classical PIR capacity is $C_{PIR}=\frac{2}{3}=\frac{30}{45}$. That is, for this case, $C_{PIR}=\frac{2}{3}<C=\frac{34}{45}$.  
\end{remark}

\begin{remark} \label{remark3}
We further expand on Remark~\ref{remark2} above by noting the following fact. The classical PIR capacity is a formula, as given in \eqref{C_PIR}, that depends only on the number of databases $N$ and the number of messages $K$, and is not necessarily achievable by the classical PIR scheme for any given message priors and lengths. To see this, we note that the classical PIR scheme requires equal message sizes. In the example in Remark~\ref{remark2} where $p_1=4p_2$ and $L_1=4L_2$, if we zero-pad the shorter message to make the message lengths the same, we achieve $R_{ach}=p_1 \frac{L_1}{D} + p_2 \frac{L_2}{D}=\frac{17}{30}$ by noting $D=\frac{3}{2} L_1$ as the length of the longer message is the common message length now, and the classical PIR capacity for this case is $\frac{2}{3}$. Thus, we observe $R_{ach}=\frac{17}{30}<C_{PIR}=\frac{2}{3}<C=\frac{34}{45}$ for this case.  
\end{remark}

As a follow up to Remark~\ref{remark3}, we note that the achievable scheme proposed in this paper always outperforms zero-padding shorter messages and applying the classical PIR scheme for so-constructed equal-length messages. This is proved in the following corollary.

\begin{corollary} \label{corollary2}
Semantic PIR capacity outperforms classical PIR rate with zero-padding.
\end{corollary}

\begin{Proof}
We first calculate the general achievable rate for the classical PIR scheme with zero-padding, $R_{ach}$. Noting $L_1 \geq L_2 \geq \cdots \geq L_K$, we zero-pad messages $2, \dots, K$ until the message sizes are all equal to $L_1$. Next, we apply the classical PIR scheme with the common message size $L_1$. Then, the download cost (and the expected download cost) becomes, 
\begin{align}
   \mathbb{E}[D]=D=\frac{L_1}{C_{PIR}} \label{R_ach_D}
\end{align}
Now, using $C_{PIR}$ in \eqref{C_PIR} in equation \eqref{R_ach_D} above, we obtain,
\begin{align}
    R_{ach}=\frac{\mathbb{E}[L]}{\mathbb{E}[D]}=\left(\frac{L_1}{\mathbb{E}[L]}+\frac{1}{N}\frac{L_1}{\mathbb{E}[L]}+\dots+\frac{1}{N^{K-1}}\frac{L_1}{\mathbb{E}[L]}\right)^{-1} \label{R_ach}
\end{align}
Note repeated $L_1$ in the expression in \eqref{R_ach}. Comparing $R_{ach}$ in \eqref{R_ach} with the semantic PIR capacity in \eqref{semantic_capacity}, we deduce that $R_{ach} \leq C$ as $L_1 \geq L_2 \geq \cdots \geq L_K$. 
\end{Proof}

\begin{remark} \label{remark4}
If all messages have equal lengths, irrespective of the prior probabilities, the capacity of semantic PIR becomes equal to that of classical PIR. Note, in this case, $L_i=\mathbb{E}[L]$ and the capacity expression in \eqref{semantic_capacity} reduces to the classical PIR capacity expression in \eqref{C_PIR}. Thus, in order to exploit variability in priors to achieve a PIR capacity higher than the classical PIR capacity, we need variability in message lengths.
\end{remark}

\section{Achievability Proof}\label{Achievability}
In this section, we present two PIR schemes that achieve the semantic PIR capacity given in Theorem~\ref{Thm2}. For each scheme, we first formally present the scheme, then we verify its correctness and privacy, calculate its achievable rate, and give explicit examples for illustration. 

\subsection{Achievable Semantic PIR Scheme 1} \label{sect:pir1}
The scheme is based on the iterative structure of the achievable scheme in \cite{sun2017capacity}. In this scheme, the user downloads $k$-sums from the messages for $k=1, \dots, K$. The novel component in our scheme is the calculation of the number of stages needed to be downloaded from each message based on the message sizes.

This achievable scheme is parameterized by $(K,N, \{L_i\}_{i=1}^K)$. Based on these parameters, the user prepares queries to retrieve the desired message privately. The basic structure of our achievable scheme is as follows.

\begin{enumerate}
  \item \textbf{Message indexing}: Order the messages in the descending order of message sizes. That is, index $1$ is assigned to the longest message and index $K$ is assigned to the shortest message ($L_1 \geq L_2 \geq \dots \geq L_K$). Calculate retrieval parameters $\upsilon_1,\upsilon_2,\dots,\upsilon_K$ corresponding to each message such that $\upsilon_1 \geq \upsilon_2 \geq \dots \geq \upsilon_K$. The retrieval parameters denote the number of stages that needs to be downloaded from each message. The explicit expressions for these parameters are given in Section~\ref{rate}. 
  
  For the rest of this section, assume that the user wishes to download $W_j$.  
  
  \item \textbf{Index preparation:} The user permutes the indices of all messages independently, uniformly, and privately from the databases.
  
  \item \textbf{Singletons:} Download $\upsilon_k$ different bits from message $W_k$ from the $n$th database, where $n=1, \dots, N$ and $k=1, \dots, K$.
  
  \item \textbf{Sums of two elements (2-sums):} There are two types of blocks in this step. The first block is the sums involving bits of the desired message, $W_j$, and the other block is the sums that do not have any bits from $W_j$. In the first block, download $(N-1)\min\{\upsilon_i,\upsilon_j\}$ bit-wise sums of $W_i$ and $W_j$ each from the $N$ databases for all $i \neq j$. Each sum comprises of an already downloaded $W_i$ bit from another database and a new bit of $W_j$. For the second block, for all possible message pairs $(W_{i_1}, W_{i_2})$ for $i_1 \neq i_2 \neq j$, download $(N-1)\min\{\upsilon_{i_1},\upsilon_{i_2}\}$ number of bit-wise sums of $W_{i_1}$ and $W_{i_2}$ each from the $N$ databases. Each sum comprises of fresh bits from $W_{i_1}$ and $W_{i_2}$. 

  \item \textbf{Repeat step 4} for all $k$-sums where $k=3, 4, \dots, K$. For each $k$-sum, download $k$ bit-wise sum from $k$ messages. If one of these messages is the desired message, the remaining $(k-1)$-sum is derived from the previous $(k-1)$th round from a different database. Otherwise, download $(N-1)^{k-1} \min\{\upsilon_{i_1}, \dots, \upsilon_{i_k}\}$ sums from new bits of the undesired messages.  
\end{enumerate}

\subsubsection{Rate of Semantic PIR Scheme 1}\label{rate}
In this PIR scheme, the total number of downloaded bits remains constant for all message requirements of the user in order to guarantee privacy. Therefore, $\mathbb{E}[D]$ in \eqref{ach-rate} can be calculated by counting the total number of bits in the set of queries sent to the databases by the user to download any message. Within the set of queries, there are $\sum_{i=1}^K N\upsilon_i$ number of singletons and $\sum_{i=t}^K  N(N-1)^{t-1}\upsilon_i\binom{i-1}{t-1}$ number of sums of $t$ elements. Therefore,
\begin{align}
    \mathbb{E}[D]&=\sum_{i=1}^K N\upsilon_i+\sum_{t=2}^K \sum_{i=t}^K  N(N-1)^{t-1}\upsilon_i\binom{i-1}{t-1}\\
    &=N\left[\sum_{i=1}^K \upsilon_i+\sum_{i=2}^K \sum_{t=2}^i (N-1)^{t-1} \upsilon_i \binom{i-1}{t-1} \right]\\
    &=N\left[\sum_{i=1}^K \upsilon_i+\sum_{i=2}^K \upsilon_i \left( \sum_{t=0}^i (N-1)^{t} \binom{i-1}{t}-1\right) \right]\\
    &=N\left[\sum_{i=1}^K \upsilon_i+\sum_{i=2}^K \upsilon_i \left( N^{i-1}-1\right) \right]\\
    &=\sum_{i=1}^K \upsilon_i N^i \label{final_D}
\end{align}

In order to calculate $\mathbb{E}[L]$, assume that the desired message is $W_j$. There are $N\upsilon_j$ number of singletons of $W_j$ in the set of queries sent to the databases to retrieve $W_j$. The scheme can recover $N(N-1)^{t-1}\upsilon_j\binom{j-1}{t-1}+N(N-1)^{t-1}\sum_{i=j+1}^K \upsilon_i\binom{i-2}{t-2}$ number of $W_j$ bits using the $t$th block of the scheme (sum of $t$ elements) when $t\leq j$, where the first term in the sum corresponds to $t$-sums with the shortest message being $W_j$ and the second term corresponds to $t$-sums with the shortest message being some other message ($\neq W_j$). When $t>j$ this scheme is able to retrieve $\sum_{i=t}^K N(N-1)^{t-1}\upsilon_i\binom{i-2}{t-2}$ number of $W_j$ bits as there should be at least $t-j$ number of messages in the sum that are shorter than $L_j$. Therefore, the total number of useful bits of $W_j$ retrieved, $U_j$, is given by,
\begin{align}
    U_j&=N\upsilon_j+\sum_{t=2}^j \left(N(N-1)^{t-1}\upsilon_j\binom{j-1}{t-1}+\sum_{i=j
    +1}^K N(N-1)^{t-1}\upsilon_i\binom{i-2}{t-2}\right)\nonumber\\
    &\quad+\sum_{t=j+1}^K\sum_{i=t}^K N(N-1)^{t-1}\upsilon_i\binom{i-2}{t-2}\\
    &=N\upsilon_j\left(1+\sum_{t=2}^j (N-1)^{t-1}\binom{j-1}{t-1}\right)+\sum_{t=2}^j\sum_{i=j+1}^K N(N-1)^{t-1}\upsilon_i\binom{i-2}{t-2}\nonumber\\
    &\quad+\sum_{t=j+1}^K\sum_{i=t}^K N(N-1)^{t-1}\upsilon_i\binom{i-2}{t-2}\\
    &=N\upsilon_j\left(1+(N-1)\binom{j-1}{1}+(N-1)^2\binom{j-1}{2}+\dots+(N-1)^{j-1}\binom{j-1}{j-1}\right)\nonumber\\
    &\quad+N\upsilon_{j+1}\left(\sum_{t=2}^j (N-1)^{t-1}\binom{j-1}{t-2}\right)+N\upsilon_{j+2}\left(\sum_{t=2}^j (N-1)^{t-1}\binom{j}{t-2}\right)+\dots\nonumber\\
    &\quad+N\upsilon_K\left(\sum_{t=2}^j (N-1)^{t-1} \binom{K-2}{t-2}\right)+N\upsilon_{j+1}(N-1)^j\binom{j-1}{j-1}\nonumber\\
    &\quad+N\upsilon_{j+2}\left((N-1)^j\binom{j}{j-1}+(N-1)^{j+1}\binom{j}{j}\right)+\dots+N\upsilon_K\left((N-1)^j\binom{K-2}{j-1}\right.\nonumber\\
    &\left. \quad+(N-1)^{j+1}\binom{K-2}{j}+\dots+(N-1)^{K-1}\binom{K-2}{K-2}\right)\\
    &=N\upsilon_j(N-1+1)^{j-1}\\
    &\quad+N\upsilon_{j+1}\left((N-1)\binom{j-1}{0}+(N-1)^2\binom{j-1}{1}+\dots+(N-1)^j\binom{j-1}{j-1}\right)\\
    &\quad+N\upsilon_{j+2}\left((N-1)\binom{j}{0}+(N-1)^2\binom{j}{1}+\dots+(N-1)^{j+1}\binom{j}{j}\right)+\dots\\
    &\quad+N\upsilon_k\left((N-1)\binom{K-2}{0}+(N-1)^2\binom{K-2}{1}+\dots+(N-1)^{K-1}\binom{K-2}{K-2}\right)\\
    &=N^j\upsilon_j+N(N-1)(N-1+1)^{j-1}\upsilon_{j+1}+N(N-1)(N-1+1)^j\upsilon_{j+2}+\dots\nonumber\\
    &\quad +N(N-1)(N-1+1)^{K-2}\upsilon_K\\
    &=N^j\upsilon_j+(N-1)\sum_{i=j+1}^K N^{i-1}\upsilon_i \label{final_U_j}
\end{align}

Thus, the scheme retrieves $N^j\upsilon_j+(N-1)\sum_{i=j+1}^K N^{i-1}\upsilon_{i}$ number of useful bits of the required message at a time. Hence, we define \emph{subpacketization} for message $W_j$ as $U_j$, where 
\begin{align} \label{subpacket1}
   U_j = N^j\upsilon_j+(N-1)\sum_{i=j+1}^K N^{i-1}\upsilon_i, \quad j=1,\dots,K 
\end{align}
We then need the message sizes to be a common multiple of their own subpacketizations, 
\begin{align} \label{subpacket2}
   L_j=\alpha U_j, \quad j=1,\dots,K 
\end{align}
We note that $\alpha$ should be the same for all $j$ in \eqref{subpacket2} to guarantee privacy. 

The requirements in \eqref{subpacket1} and \eqref{subpacket2} can be written succinctly as a matrix equation,
\begin{gather}
\begin{bmatrix}
    L_1\\L_2\\L_3\\\vdots\\L_{K-1}\\L_K    
\end{bmatrix}
=
\alpha
\begin{bmatrix}
    N & N(N-1) & N^2(N-1) & \dots & N^{K-1}(N-1)\\
    0 & N^2 & N^2(N-1) &  \dots & N^{K-1}(N-1) \\
    0 & 0 & N^3 &  \dots & N^{K-1}(N-1)\\
    \vdots & \vdots & \vdots & \vdots & \vdots \\
    0 & 0 & 0 &  \dots & N^{K-1}(N-1)\\
    0 & 0 & 0 &  \dots & N^K\
\end{bmatrix}
\begin{bmatrix}
    \upsilon_1\\\upsilon_2\\\upsilon_3\\\vdots\\\upsilon_{K-1}\\\upsilon_K    
\end{bmatrix}\label{RHS}
\end{gather}
Since $L_1,\dots,L_K$ are parameters (inputs) to the scheme, the internal parameters $\upsilon_1,\dots,\upsilon_K$ can be calculated by inverting the matrix as,
\begin{gather}
\begin{bmatrix}
\upsilon_1\\\upsilon_2\\\upsilon_3\\\vdots\\\upsilon_{K-1}\\\upsilon_K 
\end{bmatrix}
=
\frac{1}{\alpha}
\begin{bmatrix}
    \frac{1}{N} & -\frac{N-1}{N^2} & -\frac{N-1}{N^3} & \dots & -\frac{N-1}{N^{(K-1)}} & -\frac{N-1}{N^{K}}\\
    0 & \frac{1}{N^{2}} & -\frac{N-1}{N^{3}} &  \dots & -\frac{N-1}{N^{K-1}} & -\frac{N-1}{N^{K}}\\
    0 & 0 & \frac{1}{N^{3}} &  \dots & -\frac{(N-1)}{N^{K-1}} & -\frac{N-1}{N^{K}}\\
    \vdots & \vdots & \vdots & \vdots & \vdots & \vdots \\
    0 & 0 & 0 &  \dots & \frac{1}{N^{K-1}} & -\frac{N-1}{N^{K}}\\
    0 & 0 & 0 &  \dots & 0 & \frac{1}{N^{K}}\\
\end{bmatrix}
\begin{bmatrix}
     L_1\\L_2\\L_3\\\vdots\\L_{K-1}\\L_K  
\end{bmatrix} \label{solution_nu}
\end{gather}
Here, $\alpha$ should be chosen to be the greatest common divisor (gcd) of the elements of the vector resulting from multiplying the matrix and the vector on the right side of \eqref{solution_nu}. This allows the shortest subpacketization levels for all messages for increased flexibility.

The total number of bits downloaded calculated in \eqref{final_D} and the number of useful bits downloaded calculated in \eqref{final_U_j} are both within one subpacketization level. This subpacketization level downloads are repeated $\alpha$ times to download the entire file; see also \eqref{subpacket2}. Thus, we calculate the achievable rate of this scheme as,
\begin{align}
    R&=\frac{\mathbb{E}[L]}{\mathbb{E}[D]}\\
    &=\frac{\sum_{i=1}^K p_iU_i}{\sum_{i=1}^K N^i\upsilon_i}\\
    &=\frac{\frac{1}{\alpha}\sum_{i=1}^K p_iL_i}{\sum_{i=1}^K \frac{1}{\alpha}N^i(N^{-i}L_i-\sum_{j=i+1}^K (N-1)N^{-j}L_j)} \label{rate-intermediate-step}\\
    &=\frac{\mathbb{E}[L]}{\sum_{i=1}^K L_i-(N-1)\sum_{i=1}^K \sum_{j=i+1}^K N^{-j}L_jN^i}\\
    &=\frac{\mathbb{E}[L]}{\sum_{i=1}^K L_i-(N-1)\left(\sum_{j=2}^K N^{-j+1}L_j+\sum_{j=3}^K N^{-j+2}L_j+\dots+N^{-1}L_K\right)  }\\
    &=\frac{\mathbb{E}[L]}{L_1+L_2\left(1-(N-1)N^{-1}\right)+\dots+L_K\left(1-(N-1)(N^{-(K-1)}+\dots+N^{-1})\right)}\\
    &=\frac{\mathbb{E}[L]}{L_1+\frac{L_2}{N}+\frac{L_3}{N^2}+\dots+\frac{L_K}{N^{K-1}}}\\
    &=\left(\frac{L_1}{\mathbb{E}[L]}+\frac{1}{N}\frac{L_2}{\mathbb{E}[L]}+\dots+\frac{1}{N^{K-1}}\frac{L_K}{\mathbb{E}[L]}\right)^{-1}
\end{align}
where \eqref{rate-intermediate-step} follows by applying \eqref{subpacket2} in the numerator and writing $\upsilon_i$ in terms of $L_j$ using \eqref{solution_nu} in the denominator. This concludes the derivation of the achievable rate.

\begin{remark}
We assume that each message has a length which is a multiple of $N^K$ to aid smooth computation of $\upsilon_1,\dots,\upsilon_K$. Note that this is consistent with \cite{sun2017capacity}.
\end{remark}

\subsubsection{Proof of Privacy}
Since $L_1\geq L_2\geq \dots \geq L_K$ we have $\upsilon_1\geq \upsilon_2\geq \dots \geq \upsilon_K$. A given database receives a set of queries for $\upsilon_1,\upsilon_2,\dots,\upsilon_K$ numbers of bits of $W_1,W_2,\dots,W_K$, respectively, as singletons and $(N-1)^{t-1}\min\{\upsilon_{i_1},\dots,\upsilon_{i_t}\}$ bit-wise $t$-sums of $W_{i_1},\dots,W_{i_t}$, for $t=2,\dots,K$. According to the query generation procedure, no bit of any message is requested from a given database more than once as a singleton or as an element of a sum. Any given database receives the exact same set of queries in type, irrespective of the desired message of the user. Similar to other known PIR schemes, the user randomly selects a permutation of the indices of the bits in the message block to be downloaded and uses these permuted indices in the set of queries, allowing no pattern to be recognized among the bits of a given message within the query set. The a posteriori probability of the user needing $W_i$ given a realization of the set of queries received by any given database is given by,
\begin{align}
    P(\theta=i|Q=q)&=\frac{P(Q=q|\theta=i)P(\theta=i)}{\sum_{j=1}^K P(Q=q|\theta=j)P(\theta=j)} \label{priv-proof}
\end{align}
According to the procedure of generating queries, we have $P(Q=q|\theta=i)=P(Q=q|\theta=j)$ for all $i$ and $j$, as for a given database, the user sends queries to download the same number of different bits (or elements of sums) from a given message, with a random permutation of its indices, irrespective of the desired message. Thus, replacing $P(Q=q|\theta=j)$ in the denominator of \eqref{priv-proof} with $P(Q=q|\theta=i)$ gives,
\begin{align}
    P(\theta=i|Q=q)&=\frac{P(Q=q|\theta=i)P(\theta=i)}{\sum_{j=1}^K P(Q=q|\theta=i)P(\theta=j)} \\
    &=P(\theta=i)
\end{align}
which ensures that this scheme is private, since it implies that $\theta$ and $Q$ are independent.

\subsection{Examples of Semantic PIR Scheme 1} \label{sect:pir1-ex}

\subsubsection{Example 1: $N=2, K=2$, $L_1=1024$ bits, $L_2=256$ bits}
First, the message indices are independently and uniformly permuted. The first and the second messages after permutations are denoted by bits $a_i$ and $b_i$, respectively.

\begin{itemize}
\item Message indexing and calculation of $\upsilon_i$: Messages are indexed such that the first message is the longer one, and the second message is the shorter one.  Below, we will give query tables for downloading $W_1$ and $W_2$. We calculate $\upsilon_1$ and $\upsilon_2$ as,
\begin{gather}
\begin{bmatrix}
\upsilon_1\\\upsilon_2\\
\end{bmatrix}
=
\frac{1}{\alpha}
\begin{bmatrix}
    \frac{1}{2} & -\frac{1}{4}\\
    0 & \frac{1}{4}\\
\end{bmatrix}
\begin{bmatrix}
     L_1\\L_2\\ 
\end{bmatrix}
\end{gather}
where $\alpha=$gcd$\{\frac{L_1}{2}-\frac{L_2}{4}, \frac{L_2}{4}\}$. By direct substitution, we get,
\begin{gather}
    \begin{bmatrix}
\upsilon_1\\\upsilon_2\\
\end{bmatrix}
=
\frac{1}{\alpha}
\begin{bmatrix}
     448\\64\\  
\end{bmatrix}
\end{gather}
Hence, $\alpha=$gcd$\{448,64\}=64$. Therefore, $\upsilon_1=7$ and $\upsilon_2=1$. The subpacketization levels of $W_1$ and $W_2$ are $U_1=\frac{1024}{64}=16$ and $U_2=\frac{256}{64}=4$, respectively.

\item Singletons: Download $\upsilon_1=7$ bits of $W_1$ and $\upsilon_2=1$ bit of $W_2$ each from the two databases.

\item Sums of twos: Download $(N-1)\upsilon_2=1$ sum of $W_1$ and $W_2$ bits each from the two databases. Note that if $W_1$ is the desired message, the singletons of $W_2$ are used as a side information with new $W_1$ bits in the sum and vice versa.
\end{itemize}

Tables~\ref{table1} and \ref{table2} show the queries sent to the databases to retrieve $W_1$ and $W_2$, respectively. 

\begin{table}[ht]
\begin{center}
\begin{tabular}{ |c|c| }
\hline
 Database 1 & Database 2 \\
\hline
$a_1,\dots,a_{7}$ & $a_{8},\dots,a_{14}$\\
$b_1$ & $b_2$ \\
\hline
$a_{15}+b_{2}$ & $a_{16}+b_{1}$\\
\hline
\end{tabular}
\end{center}
\vspace*{-0.4cm}
\caption{The query table for the retrieval of $W_1$.}
\label{table1}
\end{table}

\begin{table}[ht]
\begin{center}
\begin{tabular}{ |c|c| }
\hline
 Database 1 & Database 2 \\
\hline
$a_1,\dots,a_{7}$ & $a_{8},\dots,a_{14}$\\
$b_1$ & $b_2$ \\
\hline
$a_{8}+b_{3}$ & $a_{1}+b_{4}$\\
\hline
\end{tabular}
\end{center}
\vspace*{-0.4cm}
\caption{The query table for the retrieval of $W_2$.}
\label{table2}
\end{table}

The rate achieved by this scheme when downloading $W_1$ is $R_1=\frac{16}{18}=\frac{8}{9}$, and the rate achieved by this scheme when downloading $W_2$ is $R_2=\frac{4}{18}=\frac{2}{9}$. Therefore, the average rate $R$ achieved by the scheme is,
\begin{align}
    R=\frac{\mathbb{E}[L]}{\mathbb{E}[D]}
    =\frac{p_1L_1+p_2L_2}{p_1D+p_2D}
    =p_1\frac{L_1}{D}+p_2\frac{L_2}{D}
    =p_1R_1+p_2R_2
    =\frac{8}{9}p_1+\frac{2}{9}p_2 \label{ex11}
\end{align}
This matches the capacity expression in Theorem~\ref{Thm2} as,
\begin{align}
    C=\left(\frac{L_1}{\mathbb{E}[L]}+\frac{1}{N}\frac{L_2}{\mathbb{E}[L]}\right)^{-1}
    =(1024p_1+256p_2)\left(1024+\frac{256}{2}\right)^{-1}
    =\frac{8}{9}p_1+\frac{2}{9}p_2 \label{ex111}
\end{align}
The classic PIR capacity for this case with equal priors is,
\begin{align}
    C&=\left(1+\frac{1}{N}\right)^{-1}=\left(1+\frac{1}{2}\right)^{-1}=\frac{2}{3}\label{ex12}
\end{align}
The semantic PIR capacity in \eqref{ex111} exceeds the classical PIR capacity in \eqref{ex12} when
\begin{align}
    \frac{8}{9}p_1+\frac{2}{9}p_2>\frac{2}{3}
\end{align}
which is when $p_1>\frac{2}{3}$. Consequently, when $p_1>\frac{2}{3}$, there is a strict gain from exploiting message semantics for PIR, in this case.

\begin{remark}
Although it is apparent in this example that the rate of semantic PIR is lower than the capacity of classical PIR for $p_1 < \frac{2}{3}$, as discussed in Remark~\ref{remark2} and Remark~\ref{remark3}, there is a subtle aspect that should be addressed for a fair comparison. To see this, let us take the case of uniform a priori distribution, i.e., $p_1=p_2=\frac{1}{2}$, i.e., a case where $p_1<\frac{2}{3}$. In this case, the semantic PIR capacity using \eqref{ex111} is $\frac{5}{9}$. In order to properly use the classical PIR scheme in \cite{sun2017capacity}, messages need to be of equal size. One way to do this is to zero-pad the shorter message to be of length 1024 bits as well. In this case, the actual retrieval rate is not $\frac{2}{3}$ as the actual message size of $W_2$ is much less. Specifically, the total download for this scheme is $D=\frac{L}{R}=\frac{1024}{2/3}=1536$. The actual retrieval rate for the classical PIR problem is,
\begin{align}
    R_{ach}=\frac{1/2 \times 1024+1/2 \times 256}{1536}=\frac{5}{12}<\frac{5}{9}<\frac{6}{9}
\end{align}
Thus, the actual achievable rate $R_{ach}$ is $\frac{5}{12}$, which is less than the semantic PIR capacity $\frac{5}{9}$, which is less than the classical PIR capacity $\frac{6}{9}$. Thus, even though the semantic PIR capacity is less than the classical PIR capacity, the semantic PIR capacity (which is achievable) is larger than the classical PIR rate with zero-padding as proved in Corollary~\ref{corollary2}.
\end{remark}

\subsubsection{Example 2: $N=4, K=3$, $L_1=8192$ bits, $L_2=2048$ bits, $L_3=512$ bits}
First, the message indices are independently and uniformly permuted. The first, second, and third messages after permutations are denoted by bits $a_i$, $b_i$ and $c_i$, respectively. 

\begin{itemize}
\item Message indexing and calculation of $\upsilon_i$: Messages are indexed such that the first message is the longest one, and the third message is the shortest one. Below, we will give the query table for downloading $W_2$, i.e., the medium-length message. The bits of $W_2$ are represented by $b_i$. We calculate $\upsilon_1$,  $\upsilon_2$ and $\upsilon_3$ as,
\begin{gather}
    \begin{bmatrix}
\upsilon_1\\\upsilon_2\\\upsilon_3\\
\end{bmatrix}
=
\frac{1}{\alpha}
\begin{bmatrix}
    \frac{1}{4} & -\frac{3}{16} & -\frac{3}{64}\\
    0 & \frac{1}{16} & -\frac{3}{64}\\
    0 & 0 & \frac{1}{64}\\
\end{bmatrix}
\begin{bmatrix}
     L_1\\L_2\\L_3\\  
\end{bmatrix}
\end{gather}
where $\alpha=$gcd$\{\frac{L_1}{4}-\frac{3L_2}{16}-\frac{3L_3}{64}, \frac{L_2}{16}-\frac{3L_3}{64}, \frac{L_3}{64}\}$. By direct substitution, we get,
\begin{gather}
    \begin{bmatrix}
\upsilon_1\\\upsilon_2\\\upsilon_3\\
\end{bmatrix}
=
\frac{1}{\alpha}
\begin{bmatrix}
     1640\\104\\8\\  
\end{bmatrix}
\end{gather}
Hence, $\alpha=$gcd$\{1640,104,8\}=8$. Therefore, $\upsilon_1=205$, $\upsilon_2=13$ and $\upsilon_3=1$. The subpacketization levels of $W_1$, $W_2$ and $W_3$ are $U_1=\frac{8192}{8}=1024$, $U_2=\frac{2048}{8}=256$ and $U_3=\frac{512}{8}=64$, respectively.

\item Singletons: Download $\upsilon_1=205$ bits of $W_1$, $\upsilon_2=13$ bits of $W_2$ and $\upsilon_3=1$ bits of $W_3$ each from the four databases.

\item Sums of twos: Download $(N-1)\upsilon_2=39$ sums of $W_1$ and $W_2$ and $(N-1)\upsilon_3=3$ sums of $W_2$ and $W_3$  bits each from the four databases. Use the downloaded singletons from $W_1$, $W_3$ as side information with new $W_2$ bits. Download $(N-1)\upsilon_3=3$ bit-wise sums of $W_1$ and $W_3$ each from the four databases using fresh bits of both messages.

\item Sums of threes: Download $(N-1)^2\upsilon_3=9$ bit-wise sums involving all three messages from each database utilizing the downloaded sums of $W_1$ and $W_3$ from the other databases in the previous step as side information. 
\end{itemize}

Table~\ref{table3} shows the queries sent to the databases to retrieve $W_2$. 

\begin{table}[ht]
\begin{center}
\begin{tabular}{ |c|c|c|c| }
\hline
 Database 1 & Database 2 & Database 3 & Database 4 \\
\hline
$a_1,\dots,a_{205}$ & $a_{206},\dots,a_{410}$ & $a_{411},\dots,a_{615}$ & $a_{616},\dots,a_{820}$\\
$b_1,\dots,b_{13}$ & $b_{14},\dots,b_{26}$ & $b_{27},\dots,b_{39}$ & $b_{40},\dots,b_{52}$\\
$c_1$ & $c_2$ & $c_3$ & $c_4$ \\
\hline\hline
$a_{206}+b_{53}$ & $a_{411}+b_{92}$ & $a_{616}+b_{131}$ & $a_1+b_{170}$\\
$\vdots$ & $\vdots$ & $\vdots$ & $\vdots$\\
$a_{218}+b_{65}$ & $a_{423}+b_{104}$ & $a_{628}+b_{143}$ & $a_{13}+b_{182}$\\
$a_{411}+b_{66}$ & $a_{616}+b_{105}$ & $a_1+b_{144}$ & $a_{206}+b_{183}$\\
$\vdots$ & $\vdots$ & $\vdots$ & $\vdots$\\
$a_{423}+b_{78}$ & $a_{628}+b_{117}$ & $a_{13}+b_{156}$ & $a_{218}+b_{195}$\\
$a_{616}+b_{79}$ & $a_{1}+b_{118}$ & $a_{206}+b_{157}$ & $a_{411}+b_{196}$\\
$\vdots$ & $\vdots$ & $\vdots$ & $\vdots$\\
$a_{628}+b_{91}$ & $a_{13}+b_{130}$ & $a_{218}+b_{169}$ & $a_{423}+b_{208}$\\
\hline
$b_{209}+c_{2}$ & $b_{212}+c_{3}$ & $b_{215}+c_{4}$ & $b_{218}+c_{1}$\\
$b_{210}+c_{3}$ & $b_{213}+c_{4}$ & $b_{216}+c_{1}$ & $b_{219}+c_{2}$\\
$b_{211}+c_{4}$ & $b_{214}+c_{1}$ & $b_{217}+c_{2}$ & $b_{220}+c_{3}$\\
\hline
$a_{821}+c_5$ & $a_{824}+c_8$ & $a_{827}+c_{11}$ & $a_{830}+c_{14}$\\
$a_{822}+c_6$ & $a_{825}+c_9$ & $a_{828}+c_{12}$ & $a_{831}+c_{15}$\\
$a_{823}+c_7$ & $a_{826}+c_{10}$ & $a_{829}+c_{13}$ & $a_{832}+c_{16}$\\
\hline\hline
$a_{824}+b_{221}+c_{8}$ & $a_{827}+b_{230}+c_{11}$ & $a_{830}+b_{239}+c_{14}$ & $a_{821}+b_{248}+c_5$\\
$a_{825}+b_{222}+c_{9}$ & $a_{828}+b_{231}+c_{12}$ & $a_{831}+b_{240}+c_{15}$ & $a_{822}+b_{249}+c_6$\\
$a_{826}+b_{223}+c_{10}$ & $a_{829}+b_{232}+c_{13}$ & $a_{832}+b_{241}+c_{16}$ & $a_{823}+b_{250}+c_7$\\
$a_{827}+b_{224}+c_{11}$ & $a_{830}+b_{233}+c_{14}$ & $a_{821}+b_{242}+c_{5}$ & $a_{824}+b_{251}+c_8$\\
$a_{828}+b_{225}+c_{12}$ & $a_{831}+b_{234}+c_{15}$ & $a_{822}+b_{243}+c_{6}$ & $a_{825}+b_{252}+c_9$\\
$a_{829}+b_{226}+c_{13}$ & $a_{832}+b_{235}+c_{16}$ & $a_{823}+b_{244}+c_{7}$ & $a_{826}+b_{253}+c_{10}$\\
$a_{830}+b_{227}+c_{14}$ & $a_{821}+b_{236}+c_{5}$ & $a_{824}+b_{245}+c_{8}$ & $a_{827}+b_{254}+c_{11}$\\
$a_{831}+b_{228}+c_{15}$ & $a_{822}+b_{237}+c_{6}$ & $a_{825}+b_{246}+c_{9}$ & $a_{828}+b_{255}+c_{12}$\\
$a_{832}+b_{229}+c_{16}$ & $a_{823}+b_{238}+c_{7}$ & $a_{826}+b_{247}+c_{10}$ & $a_{829}+b_{256}+c_{13}$\\
\hline
\end{tabular}
\end{center}
\vspace*{-0.4cm}
\caption{The query table for the retrieval of $W_2$.}
\label{table3}
\end{table}

The rate achieved by this scheme when downloading $W_2$ is $R_2=\frac{256}{1092}=\frac{64}{273}$, and the rates achieved when downloading $W_1$ and $W_3$ are $R_1=\frac{1024}{1092}=\frac{256}{273}$ and $R_3=\frac{64}{1092}=\frac{16}{273}$, respectively. Therefore, the average rate $R$ achieved by this scheme is,
\begin{align}
    R&=\frac{\mathbb{E}[L]}{\mathbb{E}[D]}=\frac{p_1L_1+p_2L_2+p_3L_3}{p_1D+p_2D+p_3D} 
    =p_1\frac{L_1}{D}+p_2\frac{L_2}{D}+p_3\frac{L_3}{D}
    =p_1R_1+p_2R_2+p_3R_3\\
    &=\frac{256}{273}p_1+\frac{64}{273}p_2+\frac{16}{273}p_3
\end{align}
This matches the capacity expression in Theorem~\ref{Thm2} as,
\begin{align}
    C&=\left(\frac{L_1}{\mathbb{E}[L]}+\frac{1}{N}\frac{L_2}{\mathbb{E}[L]}+\frac{1}{N^2}\frac{L_3}{\mathbb{E}[L]}\right)^{-1}\\
    &=(8192p_1+2048p_2+512p_3)\left(8192+\frac{2048}{4}+\frac{512}{4^2}\right)^{-1}\\
    &=\frac{256}{273}p_1+\frac{64}{273}p_2+\frac{16}{273}p_3 \label{ex21}
\end{align}
The classical PIR capacity for this case with equal priors is,
\begin{align}
    C&=\left(1+\frac{1}{N}+\frac{1}{N^2}\right)^{-1}=\left(1+\frac{1}{4}+\frac{1}{4^{2}}\right)^{-1}=\frac{16}{21} \label{ex22}
\end{align}
The semantic PIR capacity in \eqref{ex21} exceeds the classical PIR capacity in \eqref{ex22} when
\begin{align}
    \frac{256}{273}p_1+\frac{64}{273}p_2+\frac{16}{273}p_3 > \frac{16}{21}
\end{align}
which is equivalent to 
\begin{align}
    p_1+\frac{1}{5}p_2 > \frac{4}{5}
\end{align}

\subsection{Achievable PIR Scheme 2}
    The scheme is stochastic in the sense that the user has a list of different possible query structures and the user picks one of these structures randomly. This is unlike the previous scheme where the structure is deterministic and the randomness comes from the random permutations of indices.
    
    This scheme is developed for arbitrary number of databases and arbitrary message lengths that are multiples of $N-1$; the deterministic scheme in Sections~\ref{sect:pir1} and \ref{sect:pir1-ex} assumed message lengths that are multiples of $N^K$. The scheme can be viewed as an extension of the achievable scheme in \cite{ravi-leaky} to work with arbitrary number of databases and heterogeneous message sizes. Our scheme shares similarities with \cite{chao-tian}. However, our scheme differs in that it introduces database symmetry to the scheme. The basic structure of the achievable scheme is as follows. 
    
\begin{enumerate}
    \item \textbf{Message indexing:} Index all messages such that $L_1\geq L_2\geq \dots \geq L_K$. Divide all messages into $N-1$ blocks. Let $W_i^m$ be the $m$th block of $W_i$.
    
    For the rest of this section, assume that the user requires to download $W_j$.
    
    \item \textbf{Single blocks:} Use $N-1$ out of the $N$ databases to download each block of $W_j$ and download nothing from the remaining database. Consider all $N$ cyclic shifts of the blocks around the databases to obtain $N$ options for different queries that can be used to download $W_j$. These $N$ queries require the user to download $L_j$ bits in total, resulting in no side information.
    
    \item \textbf{Sums of two blocks/single blocks:} Choose one database to download $W_i^1$ where $i\neq j$ and download $W_j^m+W_i^1$ for $m=1,\dots,N-1$ from the remaining $N-1$ databases. Create $N$ query options in total by considering all $N$ cyclic shifts of the blocks, around the databases. Repeat the procedure for $W_i^\ell$ where $\ell=2,\dots,N-1$. There are a total of $N(N-1)\binom{K-1}{1}$ query options of this type. 
    
    \item \textbf{Sums of three blocks/sums of two blocks:} Choose one database to download $W_{i_1}^1+W_{i_2}^1$ where $i_1,i_2\neq j$ and download $W_j^m+W_{i_1}^1+W_{i_2}^1$ for $m=1,\dots,N-1$ from the remaining $N-1$ databases. Create $N$ query options in total by considering all $N$ cyclic shifts of the blocks around the databases. Repeat the procedure for $W_{i_1}^{\ell_1}+W_{i_2}^{\ell_2}$ where $\ell_1,\ell_2 \in \{2,\dots,N-1\}$. There are $N(N-1)^2\binom{K-1}{2}$ query options of this type.
    
    \item \textbf{Repeat step 4} up to sums of $K$ blocks/sums of $K-1$ blocks. 
\end{enumerate}

Once the user chooses a query to be sent to the $N$ databases, out of the $N^K$ options, each database might have to compute sums of messages with different lengths. All messages except the longest in the sum are zero-padded to the left to have equal-length blocks. Then, bit-wise sums are calculated.

Once the answers are received from the databases, the user might need to subtract messages of different lengths to recover the required message. In this case, according to the design of the scheme, the subtrahend will always be shorter than or equal to the length of the minuend. Hence, the subtraction operation in this context will not be any different than the usual operation.

\begin{remark}
Each query is chosen with probability $\frac{1}{N^K}$ as there are $\sum_{t=0}^K (N-1)^t\binom{K}{t}=N^K$ number of query options in total. Each element of the sum corresponds to the number of t-sums within the set of all possible queries that can be sent to a given database.
\end{remark}

\subsubsection{Rate of Semantic PIR Scheme 2}
In this PIR scheme, each query option is utilized by the user with a probability of $\frac{1}{N^K}$ to download any desired message. When analyzing all possible queries that can be sent to all databases, we note that they have the same entries (in a shuffled way) irrespective of the desired message. Since all query entries are equally probable to be sent to the databases, we calculate $\mathbb{E}[D]$ by,
\begin{align}
    \mathbb{E}[D]&=\sum_{i=1}^K p_i\frac{1}{N^K}\left(\sum_{t=1}^K \sum_{j=1}^{K-t+1} L_j(N-1)^{t-1} \binom{K-j}{t-1}\right)N \label{tsum1}\\
    &=\frac{1}{N^{K-1}} \sum_{j=1}^K \sum_{t=1}^{K-j+1}   L_j(N-1)^{t-1} \binom{K-j}{t-1} \label{tsum2}\\
    &=\frac{1}{N^{K-1}} \sum_{j=1}^K L_j \sum_{t=0}^{K-j} (N-1)^{t} \binom{K-j}{t}\\
    &=\frac{1}{N^{K-1}} \sum_{j=1}^K L_j N^{K-j}\\
    &=L_1+\frac{L_2}{N}+\frac{L_3}{N^2}+\dots+\frac{L_K}{N^{K-1}} \label{pir2-ach1}
\end{align}
where the second and third sums in \eqref{tsum1} correspond to different $t$-sums and all possible longest messages within the $t$-sum, respectively. The $p_i$ terms are ignored in \eqref{tsum2} as the expected number of downloads per query set does not depend on the desired message.

For a given desired message, the number of downloaded useful bits is the length of the desired message (ignoring zero-padding, as it is ignored by the user upon receiving the answer strings). This remains constant regardless of the query set utilized by the user. Hence, 
\begin{align}
    \mathbb{E}[L]=\sum_{i=1}^K p_iL_i \label{pir2-ach2}
\end{align}
Thus, combining \eqref{pir2-ach1} and \eqref{pir2-ach2}, the achievable rate of this scheme becomes,
\begin{align}
    R&=\frac{\mathbb{E}[L]}{\mathbb{E}[D]}\\
    &=\frac{\mathbb{E}[L]}{L_1+\frac{L_2}{N}+\frac{L_3}{N^2}\dots+\frac{L_K}{N^{K-1}}}\\
    &=\left(\frac{L_1}{\mathbb{E}[L]}+\frac{1}{N}\frac{L_2}{\mathbb{E}[L]}+\dots+\frac{1}{N^{K-1}}\frac{L_K}{\mathbb{E}[L]}\right)^{-1}
\end{align}
This concludes the derivation of the achievable rate.

\subsubsection{Proof of Privacy}
Irrespective of the desired message, the user sends one of the same set of queries to a given database with probability $\frac{1}{N^K}$. Therefore, from a given database's perspective, the a posteriori probability of the user needing message $j$, upon receiving a query $q$ from a user can be calculated by,  
\begin{align}
    P(\theta=i|Q=q)&=\frac{P(Q=q|\theta=i)P(\theta=i)}{\sum_{j=1}^K P(Q=q|\theta=j)P(\theta=j)}\\
    &=\frac{\frac{1}{N^K}P(\theta=i)}{\sum_{j=1}^K \frac{1}{N^K}P(\theta=j)}\\
    &=P(\theta=i)
\end{align}
which ensures that this scheme is private, since it implies that $\theta$ and $Q$ are independent.

\subsection{Example of Semantic PIR Scheme 2 }

\subsubsection{Example 3: $N=4, K=2$, $L_1=3000$ bits, $L_2=1800$ bits}
Tables~\ref{table4} and \ref{table5} show the sets of queries that the user can utilize with probability $\frac{1}{16}$ in order to retrieve messages $W_1$ and $W_2$, respectively. We note that for a given database, the set of possible queries that the user utilizes is the same regardless of the desired message. Whenever a set of queries for the four databases is chosen with probability $\frac{1}{16}$, the required message is retrieved by subtracting the smaller sum from the larger sums, guaranteeing correctness.

In the first block of Table~\ref{table4}, $W_1$ is divided into 3 parts and each part is retrieved from different 3 databases at each query option. In the second block, $W_2^1$ is used as side information, which is requested from one database, and the three parts of $W_1$ are retrieved from the other three databases in terms of $W_1^i+W_2^1$ for $i=1,2,3$. The third and fourth blocks are the same as block 2, with $W_2^1$ replaced by $W_2^2$ and $W_2^3$, respectively. The same procedure is carried out in Table~\ref{table5}, when the user needs to retrieve $W_2$.

\begin{table}[ht]
\begin{center}
\begin{tabular}{ |c|c|c|c|c| }
\hline
 Probability &  Database 1 & Database 2 & Database 3 & Database 4 \\
\hline
$\frac{1}{16}$ & $W_1^1$ & $W_1^2$ & $W_1^3$ & $\phi$\\
$\frac{1}{16}$ & $W_1^2$ & $W_1^3$ & $\phi$ & $W_1^1$\\
$\frac{1}{16}$ & $W_1^3$ & $\phi$ & $W_1^1$ & $W_1^2$\\
$\frac{1}{16}$ & $\phi$ & $W_1^1$ & $W_1^2$ & $W_1^3$\\
\hline
$\frac{1}{16}$ & $W_1^1+W_2^1$ & $W_1^2+W_2^1$ & $W_1^3+W_2^1$ & $W_2^1$\\
$\frac{1}{16}$ & $W_1^2+W_2^1$ & $W_1^3+W_2^1$ & $W_2^1$ & $W_1^1+W_2^1$\\
$\frac{1}{16}$ & $W_1^3+W_2^1$ & $W_2^1$ & $W_1^1+W_2^1$ & $W_1^2+W_2^1$\\
$\frac{1}{16}$ & $W_2^1$ & $W_1^1+W_2^1$ & $W_1^2+W_2^1$ & $W_1^3+W_2^1$\\
\hline
$\frac{1}{16}$ & $W_1^1+W_2^2$ & $W_1^2+W_2^2$ & $W_1^3+W_2^2$ & $W_2^2$\\
$\frac{1}{16}$ & $W_1^2+W_2^2$ & $W_1^3+W_2^2$ & $W_2^2$ & $W_1^1+W_2^2$\\
$\frac{1}{16}$ & $W_1^3+W_2^2$ & $W_2^2$ & $W_1^1+W_2^2$ & $W_1^2+W_2^2$\\
$\frac{1}{16}$ & $W_2^2$ & $W_1^1+W_2^2$ & $W_1^2+W_2^2$ & $W_1^3+W_2^2$\\
\hline
$\frac{1}{16}$ & $W_1^1+W_2^3$ & $W_1^2+W_2^3$ & $W_1^3+W_2^3$ & $W_2^3$\\
$\frac{1}{16}$ & $W_1^2+W_2^3$ & $W_1^3+W_2^3$ & $W_2^3$ & $W_1^1+W_2^3$\\
$\frac{1}{16}$ & $W_1^3+W_2^3$ & $W_2^3$ & $W_1^1+W_2^3$ & $W_1^2+W_2^3$\\
$\frac{1}{16}$ & $W_2^3$ & $W_1^1+W_2^3$ & $W_1^2+W_2^3$ & $W_1^3+W_2^3$\\
\hline
\end{tabular}
\end{center}
\vspace*{-0.4cm}
\caption{The query table for the retrieval of $W_1$.}
\label{table4}
\end{table}

\begin{table}[ht]
\begin{center}
\begin{tabular}{ |c|c|c|c|c| }
\hline
 Probability &  Database 1 & Database 2 & Database 3 & Database 4 \\
\hline
$\frac{1}{16}$ & $W_2^1$ & $W_2^2$ & $W_2^3$ & $\phi$\\
$\frac{1}{16}$ & $W_2^2$ & $W_2^3$ & $\phi$ & $W_2^1$\\
$\frac{1}{16}$ & $W_2^3$ & $\phi$ & $W_2^1$ & $W_2^2$\\
$\frac{1}{16}$ & $\phi$ & $W_2^1$ & $W_2^2$ & $W_2^3$\\
\hline
$\frac{1}{16}$ & $W_2^1+W_1^1$ & $W_2^2+W_1^1$ & $W_2^3+W_1^1$ & $W_1^1$\\
$\frac{1}{16}$ & $W_2^2+W_1^1$ & $W_2^3+W_1^1$ & $W_1^1$ & $W_2^1+W_1^1$\\
$\frac{1}{16}$ & $W_2^3+W_1^1$ & $W_1^1$ & $W_2^1+W_1^1$ & $W_2^2+W_1^1$\\
$\frac{1}{16}$ & $W_1^1$ & $W_2^1+W_1^1$ & $W_2^2+W_1^1$ & $W_2^3+W_1^1$\\
\hline
$\frac{1}{16}$ & $W_2^1+W_1^2$ & $W_2^2+W_1^2$ & $W_2^3+W_1^2$ & $W_1^2$\\
$\frac{1}{16}$ & $W_2^2+W_1^2$ & $W_2^3+W_1^2$ & $W_1^2$ & $W_2^1+W_1^2$\\
$\frac{1}{16}$ & $W_2^3+W_1^2$ & $W_1^2$ & $W_2^1+W_1^2$ & $W_2^2+W_1^2$\\
$\frac{1}{16}$ & $W_1^2$ & $W_2^1+W_1^2$ & $W_2^2+W_1^2$ & $W_2^3+W_1^2$\\
\hline
$\frac{1}{16}$ & $W_2^1+W_1^3$ & $W_2^2+W_1^3$ & $W_2^3+W_1^3$ & $W_1^3$\\
$\frac{1}{16}$ & $W_2^2+W_1^3$ & $W_2^3+W_1^3$ & $W_1^3$ & $W_2^1+W_1^3$\\
$\frac{1}{16}$ & $W_2^3+W_1^3$ & $W_1^3$ & $W_2^1+W_1^3$ & $W_2^2+W_1^3$\\
$\frac{1}{16}$ & $W_1^3$ & $W_2^1+W_1^3$ & $W_2^2+W_1^3$ & $W_2^3+W_1^3$\\
\hline
\end{tabular}
\end{center}
\vspace*{-0.4cm}
\caption{The query table for the retrieval of $W_2$.}
\label{table5}
\end{table}

The rate achieved by this scheme when retrieving $W_1$ is,
\begin{align}
    R_1&=
    \frac{L_1}{\frac{1}{16}\left(4L_1+12(\frac{L_1}{3}\times3+\frac{L_2}{3})\right)}
    =\frac{L_1}{\frac{1}{16}(16L_1+4L_2)}
    =\frac{3000}{\frac{1}{16}(16\times3000+4\times 1800)}\\
    &=\frac{20}{23}
\end{align}
The rate achieved by this scheme when retrieving $W_2$ is,
\begin{align}
    R_2&=\frac{L_2}{\frac{1}{16}\left(4L_2+12\times4\times\frac{L_1}{3}\right)}
    =\frac{L_2}{\frac{1}{16}(16L_1+4L_2)}
    =\frac{1800}{\frac{1}{16}(16\times3000+4\times1800)}\\
    &=\frac{12}{23}
\end{align}
The overall message retrieval rate for this example is,
\begin{align}
    R&=\frac{\mathbb{E}[L]}{\mathbb{E}[D]}
    =\frac{p_1L_1+p_2L_2}{p_1 D + p_2 D}=p_1 \frac{L_1}{D} + p_2 \frac{L_2}{D}
    =p_1R_1+p_2R_2
    =\frac{20}{23}p_1+\frac{12}{23}p_2
\end{align}
This matches the semantic PIR capacity expression in Theorem~\ref{Thm2},
\begin{align}
    C&=\left(\frac{L_1}{\mathbb{E}[L]}+\frac{1}{N}\frac{L_2}{\mathbb{E}[L]}\right)^{-1}
    =(3000p_1+1800p_2)\left(3000+\frac{1800}{4}\right)^{-1}
    =\frac{20}{23}p_1+\frac{12}{23}p_2 \label{ex31}
\end{align}
The classical PIR capacity for this case with equal priors is,
\begin{align}
    C=\left(1+\frac{1}{N}\right)^{-1}=\left(1+\frac{1}{4}\right)^{-1}=\frac{4}{5} \label{ex32}
\end{align}
The semantic PIR capacity in \eqref{ex31} exceeds the classical PIR capacity in \eqref{ex32} when
\begin{align}
   \frac{20}{23}p_1+\frac{12}{23}p_2>\frac{4}{5}
\end{align}
which is when $p_1>\frac{4}{5}$. Consequently, when $p_1>\frac{4}{5}$, there is a strict gain from exploiting message semantics for PIR, in this case.

\begin{remark}
We note again that the rate calculation presented here for the semantic PIR capacity takes into consideration the zero-padding needed to be added to the shorter message block in order to perform bit-wise message addition for any query realization. The classical PIR capacity expression in \eqref{ex32} assumes that all messages are of equal size and hence the extra zero-padding is not reflected in that expression. Hence, the actual rate of classical PIR scheme is indeed less than the reported PIR capacity if the messages are of unequal size.
\end{remark}

\subsubsection{Example 4: $N=3, K=3$, $L_1=400$ bits, $L_2=300$ bits and $L_3=100$ bits}

Table 6 shows the query options that the user may use with probability $\frac{1}{27}$, to download $W_1$. Whenever a set of queries for the three databases is chosen with probability $\frac{1}{27}$, the required message is retrieved by subtracting the smaller sum from the larger sums, guaranteeing correctness.

The queries in the first block have zero side information, and retrieve the $N-1=2$ parts of $W_1$ using $N-1$ different databases. The second block uses $W_2^1$ as side information, and retrieve the two parts of $W_1$ (in terms of a sum of itself and side information) using the other two databases. The same procedure is carried out in blocks 3, 4 and 5, with $W_2^1$ replaced by $W_2^2$, $W_3^1$ and $W_3^2$. Last four blocks of Table~\ref{table6} use $W_2^i+W_3^j$ for $j\in{1,2}$ as side information and use sums of three elements ($W_1^k+W_2^i+W_3^j$ for $k=1,2$) to retrieve the two parts of $W_1$. 

\begin{table}[ht]
\begin{center}
\begin{tabular}{ |c|c|c|c|c| }
\hline
 Probability &  Database 1 & Database 2 & Database 3 \\
\hline
$\frac{1}{27}$ & $W_1^1$ & $W_1^2$ & $\phi$\\
$\frac{1}{27}$ & $W_1^2$ & $\phi$ & $W_1^1$\\
$\frac{1}{27}$ & $\phi$ & $W_1^1$ & $W_1^2$\\
\hline
$\frac{1}{27}$ & $W_1^1+W_2^1$ & $W_1^2+W_2^1$ & $W_2^1$\\
$\frac{1}{27}$ & $W_1^2+W_2^1$ & $W_2^1$ & $W_1^1+W_2^1$\\
$\frac{1}{27}$ & $W_2^1$ & $W_1^1+W_2^1$ & $W_1^2+W_2^1$\\
\hline
$\frac{1}{27}$ & $W_1^1+W_2^2$ & $W_1^2+W_2^2$ & $W_2^2$\\
$\frac{1}{27}$ & $W_1^2+W_2^2$ & $W_2^2$ & $W_1^1+W_2^2$\\
$\frac{1}{27}$ & $W_2^2$ & $W_1^1+W_2^2$ & $W_1^2+W_2^2$\\
\hline
$\frac{1}{27}$ & $W_1^1+W_3^1$ & $W_1^2+W_3^1$ & $W_3^1$\\
$\frac{1}{27}$ & $W_1^2+W_3^1$ & $W_3^1$ & $W_1^1+W_3^1$\\
$\frac{1}{27}$ & $W_3^1$ & $W_1^1+W_3^1$ & $W_1^2+W_3^1$\\
\hline
$\frac{1}{27}$ & $W_1^1+W_3^2$ & $W_1^2+W_3^2$ & $W_3^2$\\
$\frac{1}{27}$ & $W_1^2+W_3^2$ & $W_3^2$ & $W_1^1+W_3^2$\\
$\frac{1}{27}$ & $W_3^2$ & $W_1^1+W_3^2$ & $W_1^2+W_3^2$\\
\hline
$\frac{1}{27}$ & $W_1^1+W_2^1+W_3^1$ & $W_1^2+W_2^1+W_3^1$ & $W_2^1+W_3^1$\\
$\frac{1}{27}$ & $W_1^2+W_2^1+W_3^1$ & $W_2^1+W_3^1$ & $W_1^1+W_2^1+W_3^1$\\
$\frac{1}{27}$ & $W_2^1+W_3^1$ & $W_1^1+W_2^1+W_3^1$ & $W_1^2+W_2^1+W_3^1$\\
\hline
$\frac{1}{27}$ & $W_1^1+W_2^2+W_3^1$ & $W_1^2+W_2^2+W_3^1$ & $W_2^2+W_3^1$\\
$\frac{1}{27}$ & $W_1^2+W_2^2+W_3^1$ & $W_2^2+W_3^1$ & $W_1^1+W_2^2+W_3^1$\\
$\frac{1}{27}$ & $W_2^2+W_3^1$ & $W_1^1+W_2^2+W_3^1$ & $W_1^2+W_2^2+W_3^1$\\
\hline
$\frac{1}{27}$ & $W_1^1+W_2^1+W_3^2$ & $W_1^2+W_2^1+W_3^2$ & $W_2^1+W_3^2$\\
$\frac{1}{27}$ & $W_1^2+W_2^1+W_3^2$ & $W_2^1+W_3^2$ & $W_1^1+W_2^1+W_3^2$\\
$\frac{1}{27}$ & $W_2^1+W_3^2$ & $W_1^1+W_2^1+W_3^2$ & $W_1^2+W_2^1+W_3^2$\\
\hline
$\frac{1}{27}$ & $W_1^1+W_2^2+W_3^2$ & $W_1^2+W_2^2+W_3^2$ & $W_2^2+W_3^2$\\
$\frac{1}{27}$ & $W_1^2+W_2^2+W_3^2$ & $W_2^2+W_3^2$ & $W_1^1+W_2^2+W_3^2$\\
$\frac{1}{27}$ & $W_2^2+W_3^2$ & $W_1^1+W_2^2+W_3^2$ & $W_1^2+W_2^2+W_3^2$\\
\hline
\end{tabular}
\end{center}
\vspace*{-0.4cm}
\caption{The query table for the retrieval of $W_1$.}
\label{table6}
\end{table}

The rate achieved by this scheme when retrieving $W_1$ is,
\begin{align}
    R_1&=
    \frac{L_1}{\frac{1}{27}\left(3L_1+18(\frac{L_1}{2}\times2+\frac{L_2}{2})+6(\frac{L_1}{2}\times2+\frac{L_3}{2})\right)}
    =\frac{L_1}{\frac{1}{27}(27L_1+9L_2+3L_3)}\\
    &=\frac{400}{\frac{1}{27}(27\times400+9\times 300+3\times100)}
    =\frac{36}{46}
\end{align}
The rate achieved by this scheme when retrieving $W_2$ is,
\begin{align}
    R_2&=\frac{L_2}{\frac{1}{27}\left(3L_2+18\times3\times\frac{L_1}{2}+6\times(L_2+\frac{L_3}{2})\right)}
    =\frac{L_2}{\frac{1}{27}(27L_1+9L_2+3L_3)}\\
    &=\frac{300}{\frac{1}{27}(27\times400+9\times300+3\times100)}
    =\frac{27}{46}
\end{align}
The rate achieved by this scheme when retrieving $W_3$ is,
\begin{align}
    R_3&=\frac{L_3}{\frac{1}{27}\left(3L_3+18\times3\times\frac{L_1}{2}+6\times3\times\frac{L_2}{2}\right)}
    =\frac{L_3}{\frac{1}{27}(27L_1+9L_2+3L_3)}\\
    &=\frac{100}{\frac{1}{27}(27\times400+9\times300+3\times100)}
    =\frac{9}{46}
\end{align}
The overall message retrieval rate for this example is,
\begin{align}
    R&=\frac{\mathbb{E}[L]}{\mathbb{E}[D]}
    =p_1 \frac{L_1}{D} + p_2 \frac{L_2}{D} + p_3 \frac{L_3}{D}
    =p_1R_1+p_2R_2+p_3R_3
    =\frac{36}{46}p_1+\frac{27}{46}p_2+\frac{9}{46}p_3
\end{align}
This matches the semantic PIR capacity expression in Theorem~\ref{Thm2},
\begin{align}
    C&=\left(\frac{L_1}{\mathbb{E}[L]}+\frac{1}{N}\frac{L_2}{\mathbb{E}[L]}+\frac{1}{N^2}\frac{L_3}{\mathbb{E}[L]}\right)^{-1}
    =(400p_1+300p_2+100p_3)\left(400+\frac{300}{3}+\frac{100}{9}\right)^{-1}\\
    &=\frac{36}{46}p_1+\frac{27}{46}p_2+\frac{9}{46}p_3 \label{ex41}
\end{align}
The classical PIR capacity for this case with equal priors is,
\begin{align}
    C=\left(1+\frac{1}{N}+\frac{1}{N^2}\right)^{-1}=\left(1+\frac{1}{3}+\frac{1}{9}\right)^{-1}=\frac{9}{13} \label{ex42}
\end{align}
The semantic PIR capacity in \eqref{ex41} exceeds the classical PIR capacity in \eqref{ex42} when
\begin{align}
   \frac{36}{46}p_1+\frac{27}{46}p_2+\frac{9}{46}p_3>\frac{9}{13}
\end{align}
which is equivalent to 
\begin{align}
  p_1+\frac{2}{3}p_2>\frac{11}{13}
\end{align}

\begin{remark}
The second scheme presented above is an extension to more than two databases of the path-based scheme presented in \cite{ravi-leaky}. It is also similar to the scheme provided in \cite{chao-tian}, except for the fact that the above scheme has database symmetry as opposed to the scheme presented in \cite{chao-tian}.
\end{remark}

\section{Converse Proof}\label{converse}
In this section, we present the converse proof for Theorem~\ref{Thm2}. We note that our converse proof inherits most of its core ideas from the original work of \cite{sun2017capacity}. The central intuition of our proof is the fact that lengths of all the answer strings should be equal as a consequence of the privacy constraint. That is, since the privacy constraint requires,
\begin{align}
    A_n^{[i]} \sim A_n^{[j]}, \quad n \in [N], \quad i,j \in [K]
\end{align}
we have $H(A_n^{[i]})=H(A_n^{[j]})$ for all $i, \, j \in [K]$ for all $n \in [N]$. The major difference of our proof compared to \cite{sun2017capacity} is the handling of the non-equal message sizes.

We begin the proof of Theorem~\ref{Thm2} by the definition of message retrieval rate,
\begin{align}
    R & = \dfrac{\mathbb{E}[L]}{\mathbb{E}[D]} \label{conv0}
\end{align}
We choose some permutation $\{i_1, \dots, i_K\}$ as an arbitrary order of the messages. The denominator of \eqref{conv0} can be expanded as follows,
\begin{align}
 \mathbb{E}[D]&=\sum_{i=1}^K q_i(H(A^{[i]}_1)+\dots+H(A^{[i]}_N))\\ 
 &= H(A^{[i_1]}_1)+\dots+H(A^{[i_1]}_N) \label{conv1}\\
 &\geq H(A^{[i_1]}_1,\dots,A^{[i_1]}_N) \label{conv2}\\
 &\geq H(A^{[i_1]}_1,\dots,A^{[i_1]}_N|Q^{[i_1]}_1,\dots,Q^{[i_1]}_N) \label{conv3}\\
 &=I(W_{i_1},\dots,W_{i_K};A^{[i_1]}_1,\dots,A^{[i_1]}_N|Q^{[i_1]}_1,\dots,Q^{[i_1]}_N) \label{conv4}\\
 &=I(W_{i_1};A^{[i_1]}_1,\dots,A^{[i_1]}_N|Q^{[i_1]}_1,\dots,Q^{[i_1]}_N)\nonumber\\
 &\quad \quad +I(W_{i_2},\dots,W_{i_K};A^{[i_1]}_1,\dots,A^{[i_1]}_N|Q^{[i_1]}_1,\dots,Q^{[i_1]}_N,W_{i_1})\\
 &= H(W_{i_1})+I(W_{i_2},\dots,W_{i_K};A^{[i_1]}_1,\dots,A^{[i_1]}_N|Q^{[i_1]}_1,\dots,Q^{[i_1]}_N,W_{i_1})\label{conv5}\\
&= L_{i_1}+I(W_{i_2},\dots,W_{i_K};A^{[i_1]}_1,\dots,A^{[i_1]}_N,Q^{[i_1]}_1,\dots,Q^{[i_1]}_N|W_{i_1})\label{conv6}\\
&\geq L_{i_1}+I(W_{i_2},\dots,W_{i_K};A^{[i_1]}_1,Q^{[i_1]}_1|W_{i_1}) \label{conv7}\\
&= L_{i_1}+I(W_{i_2},\dots,W_{i_K};A^{[i_1]}_1|Q^{[i_1]}_1,W_{i_1})\label{conv8}\\
&=L_{i_1}+H(A^{[i_1]}_1|Q^{[i_1]}_1,W_{i_1})\label{conv9}\\
&= L_{i_1}+H(A^{[i_2]}_1|Q^{[i_2]}_1,W_{i_1})\label{conv10}
\end{align}
where \eqref{conv1} follows from the privacy constraint, \eqref{conv2} follows from the independence bound, \eqref{conv3} follows from the fact that conditioning cannot increase entropy, \eqref{conv4}, \eqref{conv9} follow from the fact that the answer strings are deterministic functions of the messages and the queries, \eqref{conv5} follows from the correctness constraint and the independence of the queries and the messages, \eqref{conv6}, \eqref{conv8} follow from the independence of messages and queries, \eqref{conv7} follows from the non-negativity of the mutual information function, and finally, \eqref{conv10} follows from the privacy constraint.

Since \eqref{conv7} holds true for any pair of query, answer string, the last inequality \eqref{conv10} is also true for any $(Q_n^{[i_2]},A_n^{[i_2]})$, hence,
\begin{align}
\mathbb{E}[D]&\geq L_{i_1}+H(A^{[i_2]}_n|Q^{[i_2]}_n,W_{i_1}), \quad n=1,\dots,N \label{conv11}
\end{align}

By summing all $N$ inequalities corresponding to \eqref{conv11} and repeating the previous arguments for $W_{i_2}$ (with conditioning on $W_{i_1}$) leads to,
\begin{align}
N\mathbb{E}[D]&\geq NL_{i_1}+H(A^{[i_2]}_1|Q^{[i_2]}_1,W_{i_1})+\dots+H(A^{[i_2]}_N|Q^{[i_2]}_N,W_{i_1})\\
&\geq NL_{i_1}+H(A^{[i_2]}_1,\dots,A^{[i_2]}_N|Q^{[i_2]}_1,\dots,Q^{[i_2]}_N,W_{i_1}) \label{conv12}\\
&=NL_{i_1}+I(W_{i_2},\dots,W_{i_K};A^{[i_2]}_1,\dots,A^{[i_2]}_N|Q^{[i_2]}_1,\dots,Q^{[i_2]}_N,W_{i_1})\\
&=NL_{i_1}+I(W_{i_2};A^{[i_2]}_1,\dots,A^{[i_2]}_N|Q^{[i_2]}_1,\dots,Q^{[i_2]}_N,W_{i_1})\nonumber\\
&\quad +I(W_{i_3},\dots,W_{i_K};A^{[i_2]}_1,\dots,A^{[i_2]}_N|Q^{[i_2]}_1,\dots,Q^{[i_2]}_N,W_{i_1},W_{i_2})\\
&= NL_{i_1}+L_{i_2}+I(W_{i_3},\dots,W_{i_K};A^{[i_2]}_1,\dots,A^{[i_2]}_N,Q^{[i_2]}_1,\dots,Q^{[i_2]}_N|W_{i_1},W_{i_2})\\
&\geq NL_{i_1}+L_{i_2}+I(W_{i_3},\dots,W_{i_K};A^{[i_2]}_1,Q^{[i_2]}_1|W_{i_1},W_{i_2})\\
&= NL_{i_1}+L_{i_2}+I(W_{i_3},\dots,W_{i_K};A^{[i_2]}_1|Q^{[i_2]}_1,W_{i_1},W_{i_2})\\
&=NL_{i_1}+L_{i_2}+H(A^{[i_2]}_1|Q^{[i_2]}_1,W_{i_1},W_{i_2})\\
&\geq NL_{i_1}+L_{i_2}+H(A^{[i_3]}_1|Q^{[i_3]}_1,W_{i_1},W_{i_2})
\end{align}

The last inequality holds for all $(Q_n^{[i_2]},A_n^{[i_2]}, \: n \in [N])$, hence,
\begin{align}
N\mathbb{E}[D]&\geq NL_{i_1}+L_{i_2}+I(W_{i_3},\dots,W_{i_K};A^{[i_2]}_n,Q^{[i_2]}_n|W_{i_1},W_{i_2}),\quad n=1,\dots,N\\
&\geq NL_{i_1}+L_{i_2}+H(A^{[i_3]}_n|Q^{[i_3]}_n,W_{i_1},W_{i_2}),\quad n=1,\dots,N
\end{align}

By summing the corresponding inequalities and continuing with the same procedure for $W_{i_3}, \dots, W_{i_K}$ as we have done for $W_{i_1}, W_{i_2}$, we have,
\begin{align}
N^{K-1}\mathbb{E}[D] &\geq N^{K-1}L_{i_1}+N^{K-2}L_{i_2}+\dots+NL_{i_{K-1}}\nonumber\\
&\quad+I(W_{i_K};A^{[i_K]}_1,\dots+A^{[i_K]}_N|Q^{[i_K]}_1,\dots,Q^{[i_K]}_N,W_{i_1},\dots,W_{i_{K-1}})
\end{align}
and therefore, we have,
\begin{align}
\mathbb{E}[D]&\geq L_{i_1}+\frac{1}{N}L_{i_2}+\dots+\frac{1}{N^{K-2}}L_{i_{K-1}}+\frac{1}{N^{K-1}}L_{i_K}
\end{align}
This gives,
\begin{align}
\frac{\mathbb{E}[L]}{\mathbb{E}[D]}&\leq \frac{\mathbb{E}[L]}{L_{i_1}+\frac{1}{N}L_{i_2}+\dots+\frac{1}{N^{K-2}}L_{i_{K-1}}+\frac{1}{N^{K-1}}L_{i_K}}
\end{align}
which further gives,
\begin{align}
R&\leq \left(\frac{L_{i_1}}{\mathbb{E}[L]}+\frac{1}{N}\frac{L_{i_2}}{\mathbb{E}[L]}+\dots+\frac{1}{N^{K-1}}\frac{L_{i_K}}{\mathbb{E}[L]}\right)^{-1} \label{final-ub}
\end{align}

The upper bound in \eqref{final-ub} holds for any permutation $\{i_1, \dots, i_K\}$, hence, the tightest upper bound can be obtained by minimizing over all permutations\footnote{Note that the order does not matter in the case of equal message lengths in \cite{sun2017capacity}.}. Consequently, 
\begin{align}
  R&\leq \min_{\{i_1, \dots, i_K\}} \left(\frac{L_{i_1}}{\mathbb{E}[L]}+\frac{1}{N}\frac{L_{i_2}}{\mathbb{E}[L]}+\dots+\frac{1}{N^{K-1}}\frac{L_{i_K}}{\mathbb{E}[L]}\right)^{-1}  
\end{align}
Since the messages are ordered such that $L_1 \geq L_2 \geq \dots \geq L_K$, the minimum upper bound is attained at $\{i_1, \dots, i_K\}=\{1, \dots, K\}$ as it gives the largest number to the largest coefficient in the lower bound on the download cost. Thus,
\begin{align}
    R \leq \left(\frac{L_{1}}{\mathbb{E}[L]}+\frac{1}{N}\frac{L_{2}}{\mathbb{E}[L]}+\dots+\frac{1}{N^{K-1}}\frac{L_{K}}{\mathbb{E}[L]}\right)^{-1} 
\end{align}
completing the converse proof.

\section{Conclusion and Discussion}
In this work, we introduced the problem of semantic PIR. In this problem, the stored messages are allowed to have non-uniform popularities, which is captured via an a priori probability distribution $(p_i, \: i \in [K])$, and heterogeneous sizes $(L_i, \: i \in [K])$. We derived the exact semantic PIR capacity as a function of $\{L_i\}_{i=1}^K$ and the expected message size $\mathbb{E}[L]$. The result implies that the semantic PIR capacity is equal to the classical PIR capacity if all messages have equal sizes $L_i=L$ for all $i \in [K]$. We derived a necessary and sufficient condition for the semantic PIR capacity to exceed the classical PIR capacity. In particular, we showed that if the longer messages are retrieved more often, there is a strict retrieval rate gain from exploiting the message semantics. Furthermore, we proved that for all message sizes and priors, the semantic PIR capacity exceeds the achievable rate of classical PIR with zero-padding, which zero-pads all messages to equalize their sizes.

To that end, we proposed two achievable schemes for achieving the semantic PIR capacity. The first one has a deterministic query structure. We have proposed a systematic way of calculating the needed subpacketization levels for the messages. The second scheme has a stochastic query structure, where the user picks one query structure at random from an ensemble of structures. The first scheme has the advantage of having a fixed download cost for all messages for all query structures unlike the stochastic scheme, which has the same expected download cost. Nevertheless, the first scheme suffers from exponential subpacketization levels in contrast to the linear counterpart in the stochastic scheme. Finally, we derived a matching converse that extends the converse scheme of \cite{sun2017capacity} to take into account the heterogeneous message sizes and prior probabilities. 

\bibliographystyle{unsrt}
\bibliography{reference}

\begin{thebibliography}{10}

\bibitem{PIR_ORI}
B.~Chor, E.~Kushilevitz, O.~Goldreich, and M.~Sudan.
\newblock Private information retrieval.
\newblock {\em Journal of the ACM}, 45(6):965--981, November 1998.

\bibitem{yekhanin2010private}
S.~Yekhanin.
\newblock Private information retrieval.
\newblock {\em Communications of the ACM}, 53(4):68--73, April 2010.

\bibitem{sun2017capacity}
H.~Sun and S.~A. Jafar.
\newblock The capacity of private information retrieval.
\newblock {\em IEEE Trans. on Info. Theory}, 63(7):4075--4088, July 2017.

\bibitem{JafarColluding}
H.~Sun and S.~A. Jafar.
\newblock The capacity of robust private information retrieval with colluding
  databases.
\newblock {\em IEEE Trans. on Info. Theory}, 64(4):2361--2370, April 2018.

\bibitem{arbitraryCollusion}
R.~Tajeddine, O.~W. Gnilke, D.~Karpuk, R.~Freij-Hollanti, C.~Hollanti, and
  S.~El Rouayheb.
\newblock Private information retrieval schemes for coded data with arbitrary
  collusion patterns.
\newblock In {\em IEEE ISIT}, June 2017.

\bibitem{disjointJafar}
Z.~{Jia}, H.~{Sun}, and S.~A. {Jafar}.
\newblock The capacity of private information retrieval with disjoint colluding
  sets.
\newblock In {\em IEEE Globecom}, December 2017.

\bibitem{arbitrary-coll-Kang}
X.~{Yao}, N.~{Liu}, and W.~{Kang}.
\newblock The capacity of private information retrieval under arbitrary
  collusion patterns.
\newblock Available at arXiv:2001.03843.

\bibitem{Staircase_PIR}
R.~Bitar and S.~El Rouayheb.
\newblock Staircase-{PIR}: Universally robust private information retrieval.
\newblock In {\em IEEE ITW}, pages 1--5, November 2018.

\bibitem{codedsymmetric}
Q.~{Wang} and M.~{Skoglund}.
\newblock Symmetric private information retrieval from mds coded distributed
  storage with non-colluding and colluding servers.
\newblock {\em IEEE Trans. on Info. Theory}, 65(8):5160--5175, August 2019.

\bibitem{SPIR_Mismatched}
Q.~Wang, H.~Sun, and M.~Skoglund.
\newblock Symmetric private information retrieval with mismatched coded
  messages and randomness.
\newblock In {\em IEEE ISIT}, pages 365--369, July 2019.

\bibitem{SPIR}
H.~Sun and S.~A. Jafar.
\newblock The capacity of symmetric private information retrieval.
\newblock {\em IEEE Transactions on Information Theory}, 65(1):322--329,
  January 2019.

\bibitem{ChaoTian_leakage}
T.~Guo, R.~Zhou, and C.~Tian.
\newblock On the information leakage in private information retrieval systems.
\newblock Available at arXiv: 1909.11605.

\bibitem{KarimCoded}
K.~Banawan and S.~Ulukus.
\newblock The capacity of private information retrieval from coded databases.
\newblock {\em IEEE Trans. on Info. Theory}, 64(3):1945--1956, March 2018.

\bibitem{codedcolluded}
R.~Freij-Hollanti, O.~Gnilke, C.~Hollanti, and D.~Karpuk.
\newblock Private information retrieval from coded databases with colluding
  servers.
\newblock {\em SIAM Journal on Applied Algebra and Geometry}, 1(1):647--664,
  2017.

\bibitem{codedcolludingZhang}
Y.~Zhang and G.~Ge.
\newblock A general private information retrieval scheme for {MDS} coded
  databases with colluding servers.
\newblock {\em Designs, Codes and Cryptography}, 87(11), November 2019.

\bibitem{Kumar_PIRarbCoded}
S.~Kumar, H.-Y. Lin, E.~Rosnes, and A.~G. i~Amat.
\newblock Achieving maximum distance separable private information retrieval
  capacity with linear codes.
\newblock {\em IEEE Trans. on Information Theory}, 65(7):4243--4273, July 2019.

\bibitem{codedcolludingJafar}
H.~Sun and S.~A. Jafar.
\newblock Private information retrieval from {MDS} coded data with colluding
  servers: Settling a conjecture by {F}reij-{H}ollanti et al.
\newblock {\em IEEE Trans. on Info. Theory}, 64(2):1000--1022, February 2018.

\bibitem{MM-PIR}
K.~Banawan and S.~Ulukus.
\newblock Multi-message private information retrieval: Capacity results and
  near-optimal schemes.
\newblock {\em IEEE Trans. on Info. Theory}, 64(10):6842--6862, October 2018.

\bibitem{MPIRcodedcolludingZhang}
Y.~Zhang and G.~Ge.
\newblock Multi-file private information retrieval from {MDS} coded databases
  with colluding servers.
\newblock Available at arXiv: 1705.03186.

\bibitem{BPIRjournal}
K.~Banawan and S.~Ulukus.
\newblock The capacity of private information retrieval from {B}yzantine and
  colluding databases.
\newblock {\em IEEE Trans. on Info. Theory}, 65(2):1206--1219, February 2019.

\bibitem{CodeColludeByzantinePIR}
R.~Tajeddine, O.~W. Gnilke, D.~Karpuk, R.~Freij-Hollanti, and C.~Hollanti.
\newblock Private information retrieval from coded storage systems with
  colluding, {B}yzantine, and unresponsive servers.
\newblock {\em IEEE Trans. on Info. Theory}, 65(6):3898--3906, June 2019.

\bibitem{Byzantine-Kang}
X.~{Yao}, N.~{Liu}, and W.~{Kang}.
\newblock The capacity of multi-round private information retrieval from
  {B}yzantine databases.
\newblock In {\em IEEE ISIT}, July 2019.

\bibitem{tandon2017capacity}
R.~Tandon.
\newblock The capacity of cache aided private information retrieval.
\newblock In {\em Allerton Conference}, October 2017.

\bibitem{KimCache}
M.~Kim, H.~Yang, and J.~Lee.
\newblock Cache-aided private information retrieval.
\newblock In {\em Asilomar Conference}, October 2017.

\bibitem{wei2017fundamental}
Y.-P. Wei, K.~Banawan, and S.~Ulukus.
\newblock Fundamental limits of cache-aided private information retrieval with
  unknown and uncoded prefetching.
\newblock {\em IEEE Trans. on Info. Theory}, 65(5):3215--3232, May 2019.

\bibitem{wei2017fundamental_partial}
Y.-P. Wei, K.~Banawan, and S.~Ulukus.
\newblock Cache-aided private information retrieval with partially known
  uncoded prefetching: Fundamental limits.
\newblock {\em IEEE JSAC}, 36(6):1126--1139, June 2018.

\bibitem{PIR_cache_edge}
S.~Kumar, A.~G. i~Amat, E.~Rosnes, and L.~Senigagliesi.
\newblock Private information retrieval from a cellular network with caching at
  the edge.
\newblock {\em IEEE Trans. on Communications}, 67(7):4900--4912, July 2019.

\bibitem{kadhe2017private}
S.~{Kadhe}, B.~{Garcia}, A.~{Heidarzadeh}, S.~{El Rouayheb}, and
  A.~{Sprintson}.
\newblock Private information retrieval with side information.
\newblock {\em IEEE Trans. on Info. Theory}, 66(4):2032--2043, April 2020.

\bibitem{chen2017capacity}
Z.~Chen, Z.~Wang, and S.~Jafar.
\newblock The capacity of ${T}$-private information retrieval with private side
  information.
\newblock Available at arXiv:1709.03022.

\bibitem{wei2017capacity}
Y.-P. Wei, K.~Banawan, and S.~Ulukus.
\newblock The capacity of private information retrieval with partially known
  private side information.
\newblock {\em IEEE Trans. on Info. Theory}, 65(12):8222--8231, December 2019.

\bibitem{MMPIR_PSI}
S.~P. Shariatpanahi, M.~J. Siavoshani, and M.~A. Maddah-Ali.
\newblock Multi-message private information retrieval with private side
  information.
\newblock In {\em IEEE ITW}, pages 1--5, November 2018.

\bibitem{SSMMPIR_SI1}
A.~Heidarzadeh, B.~Garcia, S.~Kadhe, S.~E. Rouayheb, and A.~Sprintson.
\newblock On the capacity of single-server multi-message private information
  retrieval with side information.
\newblock In {\em Allerton Conference}, pages 180--187, October 2018.

\bibitem{SSMMPIR_SI2}
S.~Li and M.~Gastpar.
\newblock Single-server multi-message private information retrieval with side
  information.
\newblock In {\em Allerton Conference}, pages 173--179, October 2018.

\bibitem{LiConverse}
S.~Li and M.~Gastpar.
\newblock Converse for multi-server single-message {PIR} with side information.
\newblock Available at arXiv:1809.09861.

\bibitem{StorageConstrainedPIR_Wei}
Y.-P. {Wei} and S.~{Ulukus}.
\newblock The capacity of private information retrieval with private side
  information under storage constraints.
\newblock {\em IEEE Trans. on Info. Theory}, 66(4):2023--2031, April 2020.

\bibitem{PrivateComputation}
H.~Sun and S.~A. Jafar.
\newblock The capacity of private computation.
\newblock {\em IEEE Trans. on Info. Theory}, 65(6):3880--3897, June 2019.

\bibitem{mirmohseni2017private}
M.~Mirmohseni and M.~A. Maddah-Ali.
\newblock Private function retrieval.
\newblock In {\em IWCIT}, pages 1--6, April 2018.

\bibitem{PrivateSearch}
Z.~Chen, Z.~Wang, and S.~Jafar.
\newblock The asymptotic capacity of private search.
\newblock In {\em IEEE ISIT}, June 2018.

\bibitem{abdul2017private}
M.~Abdul-Wahid, F.~Almoualem, D.~Kumar, and R.~Tandon.
\newblock Private information retrieval from storage constrained databases --
  coded caching meets {PIR}.
\newblock Available at arXiv:1711.05244.

\bibitem{StorageConstrainedPIR}
M.~A. Attia, D.~Kumar, and R.~Tandon.
\newblock The capacity of private information retrieval from uncoded storage
  constrained databases.
\newblock Available at arXiv:1805.04104v2.

\bibitem{PIR_decentralized}
Y.-P. Wei, B.~Arasli, K.~Banawan, and S.~Ulukus.
\newblock The capacity of private information retrieval from decentralized
  uncoded caching databases.
\newblock {\em Information}, 10, December 2019.

\bibitem{heteroPIR}
K.~Banawan, B.~Arasli, Y.-P. Wei, and S.~Ulukus.
\newblock The capacity of private information retrieval from heterogeneous
  uncoded caching databases.
\newblock {\em IEEE Trans. on Info. Theory}, 2020.
\newblock Early Access.

\bibitem{efficient_storage_ITW2019}
K.~Banawan, B.~Arasli, and S.~Ulukus.
\newblock Improved storage for efficient private information retrieval.
\newblock In {\em IEEE ITW}, August 2019.

\bibitem{Chao_storage_cost}
C.~Tian.
\newblock On the storage cost of private information retrieval.
\newblock Available at arXiv:1910.11973.

\bibitem{TamoISIT}
N.~Raviv and I.~Tamo.
\newblock Private information retrieval in graph based replication systems.
\newblock In {\em IEEE ISIT}, June 2018.

\bibitem{Karim_nonreplicated}
K.~Banawan and S.~Ulukus.
\newblock Private information retrieval from non-replicated databases.
\newblock In {\em IEEE ISIT}, pages 1272--1276, July 2019.

\bibitem{SecurePIR}
Q.~Wang and M.~Skoglund.
\newblock On {PIR} and symmetric {PIR} from colluding databases with
  adversaries and eavesdroppers.
\newblock {\em IEEE Trans. on Info. Theory}, 65(5):3183--3197, May 2019.

\bibitem{securePIRcapacity}
Q.~Wang, H.~Sun, and M.~Skoglund.
\newblock The capacity of private information retrieval with eavesdroppers.
\newblock {\em IEEE Trans. on Info. Theory}, 65(5):3198--3214, May 2019.

\bibitem{PIR_WTC_II}
K.~Banawan and S.~Ulukus.
\newblock Private information retrieval through wiretap channel {II}: Privacy
  meets security.
\newblock {\em IEEE Trans. on Info. Theory}, 2020.
\newblock Early Access.

\bibitem{securestoragePIR}
H.~Yang, W.~Shin, and J.~Lee.
\newblock Private information retrieval for secure distributed storage systems.
\newblock {\em IEEE Trans. on Info. Forensics and Security}, 13(12):2953--2964,
  December 2018.

\bibitem{XSTPIR}
Z.~Jia, H.~Sun, and S.~Jafar.
\newblock Cross subspace alignment and the asymptotic capacity of ${X}$-secure
  ${T}$-private information retrieval.
\newblock {\em IEEE Trans. on Info. Theory}, 65(9):5783--5798, September 2019.

\bibitem{arbmsgPIR}
H.~Sun and S.~A. Jafar.
\newblock Optimal download cost of private information retrieval for arbitrary
  message length.
\newblock {\em IEEE Trans. on Info. Forensics and Security}, 12(12):2920--2932,
  December 2017.

\bibitem{ChaoTian_coded_minsize}
R.~Zhou, C.~Tian, H.~Sun, and T.~Liu.
\newblock Capacity-achieving private information retrieval codes from
  {MDS}-coded databases with minimum message size.
\newblock Available at arXiv: 1903.08229.

\bibitem{MultiroundPIR}
H.~Sun and S.~A. Jafar.
\newblock Multiround private information retrieval: Capacity and storage
  overhead.
\newblock {\em IEEE Trans. on Info. Theory}, 64(8):5743--5754, August 2018.

\bibitem{KarimAsymmetricPIR}
K.~Banawan and S.~Ulukus.
\newblock Asymmetry hurts: Private information retrieval under
  asymmetric-traffic constraints.
\newblock {\em IEEE Trans. on Info. Theory}, 65(11):7628--7645, November 2019.

\bibitem{noisyPIR}
K.~Banawan and S.~Ulukus.
\newblock Noisy private information retrieval: On separability of channel
  coding and information retrieval.
\newblock {\em IEEE Trans. on Info. Theory}, 65(12):8232--8249, December 2019.

\bibitem{PIR_lifting}
R.~G.~L. D'Oliveira and S.~El Rouayheb.
\newblock One-shot {PIR}: Refinement and lifting.
\newblock {\em IEEE Trans. on Info. Theory}, 66(4):2443--2455, April 2020.

\bibitem{PIR_networks}
R.~Tajeddine, A.~Wachter-Zeh, and C.~Hollanti.
\newblock Private information retrieval over random linear networks.
\newblock Available at arXiv:1810.08941.

\bibitem{PSIjournal}
Z.~Wang, K.~Banawan, and S.~Ulukus.
\newblock Private set intersection: A multi-message symmetric private
  information retrieval perspective.
\newblock Available at arXiv: 1912.13501.

\bibitem{PIRlatent}
I.~Samy, M.~A. Attia, R.~Tandon, and L.~Lazos.
\newblock Latent-variable private information retrieval.
\newblock Available at arXiv: 2001.05998.

\bibitem{8437880}
J.~{Xu} and Z.~{Zhang}.
\newblock Building capacity-achieving {PIR} schemes with optimal
  sub-packetization over small fields.
\newblock In {\em IEEE ISIT}, pages 1749--1753, June 2018.

\bibitem{ravi-leaky}
I.~{Samy}, R.~{Tandon}, and L.~{Lazos}.
\newblock On the capacity of leaky private information retrieval.
\newblock In {\em IEEE ISIT}, pages 1262--1266, July 2019.

\bibitem{chao-tian}
C.~{Tian}, H.~{Sun}, and J.~{Chen}.
\newblock Capacity-achieving private information retrieval codes with optimal
  message size and upload cost.
\newblock {\em IEEE Trans. on Info. Theory}, 65(11):7613--7627, Nov 2019.

\end{thebibliography}
\end{document}